\documentstyle[12pt,a4]{article}
\begin{document}
\input{epsf}
\topmargin 0pt
\headheight 0pt
\headsep 0pt
\textheight 9in
\topmargin 0in
 
\def\a{\alpha}
\def\b{\beta}
\def\c{\gamma}
\def\d{\delta}
\def\e{\epsilon}
\def\h{\eta}
\def\k{\kappa}
\def\l{\lambda}
\def\m{\mu}
\def\n{\nu}
\def\o{\theta}
\def\p{\pi}
\def\r{\rho}
\def\s{\sigma}
\def\t{\tau}
\def\u{\upsilon}
\def\w{\omega}
\def\x{\chi}
\def\y{xi}
\def\z{\zeta}
 
\def\C{\Gamma}
\def\D{\Delta}
\def\L{\Lambda}
\def\O{\Theta}
\def\P{\Pi}
\def\S{\Sigma}
\def\W{\Omega}
 
\def\ux{\underline x}
\def\uy{\underline y}
\def\uz{\underline z}
\def\uu{\underline u}
\def\uv{\underline v}
\def\uk{\underline k}
\def\up{\underline p}
\def\pl{\partial}
\def\vphi{\varphi}
\def\rt{\rightarrow}

\newcommand{\schr} {Schr\"{o}dinger}
\def\be{\begin{equation}}
\def\ee{\end{equation}}
\def\bl{(\lambda)}
\def\ll{\lambda}
\def\tphistar{\widetilde{\phi}^\ast}
\def\tpsistar{\widetilde{\psi}^\ast}
\def\tphi{\widetilde{\phi}}
\def\tpsi{\widetilde{\psi}}
\def\phistar{\phi^\ast}
\def\psistar{\psi^\ast}
\def\tDelta{\widetilde{\Delta}}
\def\half{\frac{1}{2}} 
\def\tG{\widetilde{\cal G}} 
\newcommand{\sfrac}[2]{\mbox{$\frac{#1}{#2}$}}
 
\begin{titlepage}
{\hfill SWAT 96/56}

{\hfill April, 1996.}
\vspace*{2cm}

\centerline{\Large{\bf The Schr\"odinger Wave Functional}}
\vspace{0.7cm}
\centerline{\Large{\bf and Vacuum States in Curved Spacetime}}
\vspace{1.5cm}
\centerline{\bf D.V.~Long\footnote{\tt D.V.Long@swansea.ac.uk} 
\ \  and \ \ G.M.~Shore\footnote{\tt G.M.Shore@swansea.ac.uk}}
\vspace{0.8cm}
\centerline{\it Department of Physics}
\centerline{\it University of Wales, Swansea}
\centerline{\it Singleton Park}
\centerline{\it Swansea, SA2 8PP, U.K. }
\vspace{1.5cm}
\begin{abstract}
The Schr\"odinger picture description of vacuum states in curved spacetime
is further developed. General solutions for the vacuum wave functional 
are given for both static and dynamic (Bianchi type I) spacetimes
and for conformally static spacetimes of Robertson-Walker type.
The formalism is illustrated for simple cosmological models with
time-dependent metrics and the phenomenon of particle creation
is related to a special form of the kernel in the vacuum wave functional. 
\end{abstract}
\vspace{0.5cm}



\hfill
\end{titlepage}

\baselineskip=16pt plus 2pt

\section{Introduction}

Many of the unusual and potentially paradoxical phenomena
which occur in quantum field theories in curved spacetimes reflect
the nature of the vacuum state, for a review see for example
\cite{birrel82, fulling89, grib94, moss96}. 
However, whereas in Minkowski spacetime
we have a simple and unambiguous prescription for determining
the vacuum, the specification of the vacuum state in a general curved
spacetime involves many subtleties. Indeed, it is not assured that
a state satisfying the defining attributes of the Minkowski vacuum
even exists for a general spacetime.
 
In this and a companion paper \cite{me96}, we discuss the nature of the vacuum 
state for a broad class of flat and curved spacetimes and develop a 
formalism of sufficient generality to allow a future application to the 
vacuum structure of black hole spacetimes and the associated 
Hawking radiation \cite{hawking75,moss96}.
 
This formalism is based on the Schr\"odinger wave functional. The
essential reason for this choice is to avoid the conventional description
of the vacuum as the `no-particle' state in a Fock space of states
generated by the creation operators of particles defined with respect to
a particular mode decomposition of the quantum field. This approach is
inherently problematic in curved spacetime, where there is no
uniquely favoured mode decomposition and no
guarantee that the usual concept of a particle is a good description
of the spectrum of the theory.
The Schr\"odinger wave functional circumvents this second
problem by giving an intrinsic description of the vacuum without
reference to the spectrum of excited states.

The wave functional that defines the vacuum state satisfies 
a functional Schr\"o- dinger equation, which presupposes a foliation of
spacetime into a series of spacelike hypersurfaces, the Schr\"odinger
equation describing the evolution of the wave functional between
successive hypersurfaces. The wave functional therefore depends on the
choice of foliation.\footnote{Notice that the choice of foliation is quite distinct
from a choice of coordinates. The wave functional is
independent of any particular coordinates used to describe either the
spacetime or the foliation (see section 3). In the Schr\"odinger formalism, 
the choice of foliation replaces the mode decomposition in the conventional
approach. One of the advantages of this formalism is the clear separation
it provides between the choice of vacuum state and the issue of
coordinate independence.}
Our expectation is that physical quantities should be independent
of this foliation. However, intermediate quantities such as Green
functions may have a foliation dependence. This dependence should be
controlled by a functional Ward identity (somewhat analogous to the
dependence of Green functions on the choice of gauge in ordinary quantum
field theory). Physical quantities should depend only on the spacetime
manifold and boundary conditions.
 
In this paper, which is intended as the first step in a general
programme, we consider a class of globally hyperbolic spacetimes,
viz.~manifolds with the topology
\({\cal M} = {\cal R} \times \Sigma\) where \(\Sigma\)
are spacelike hypersurfaces.
Within this category, we distinguish manifolds which are
`static' and `dynamic', essentially implying that the metric depends only
on the spacelike and timelike coordinates respectively.
The class referred to as `dynamic' is in fact the Bianchi type I spacetimes.
We also consider more general conformally static spacetimes
where the conformal factor is purely time dependent.This includes the
cosmologically important class of Robertson-Walker spacetimes.

The ultimate goal of our work is a full analysis of the physics of the vacuum 
state in the Schwarzschild and Kruskal spacetimes describing collapsed
and eternal black holes. In this and the companion paper, we develop all
the necessary theoretical formalism and, as a warm-up, illustrate the 
techniques on a number of simple illustrative examples.
The solutions for static and dynamic spacetimes are sufficient to encompass
the full Kruskal metric and we check explicitly that
subtleties involving foliation dependence and the interpretation of vacuum
states are under control. The extension to black hole spacetimes
is in principle a straightforward application of these techniques, although
the presence of real event horizons separating the static and dynamic
regions is an important new feature.

A number of the results and techniques described here can be found already
elsewhere in the literature. In these cases, we have referenced the original 
works as fully as possible, but have included the material as we have been 
concerned to present a coherent and systematic study of the Schr\"odinger 
wave functional. 

\vspace{0.3cm}
The paper is organised as follows.
In section 2 we review the Schr\"odinger picture formulation of quantum 
field theory in Minkowski spacetime and derive the Schr\"odinger equation 
and its solutions for the vacuum and excited states for a free scalar field.
We also discuss some properties of a general Gaussian state which
will be needed in sections 4 and 5.
 
Then in section 3 we construct the Schr\"odinger wave functional
equation for an arbitrary foliation of a globally hyperbolic spacetime
with topology \({\cal M} = {\cal R}\times\Sigma\). This makes explicit
the dependence of the equation, and therefore the vacuum wave
functional, on the `lapse' and `shift' functions specifying the
foliation. The simplification for special coordinate choices
which mirror the foliation is shown.
We find explicit solutions for the vacuum wave functional
for the special cases of static and dynamic spacetimes
and a particular class of conformally static metrics including the
Robertson-Walker spacetimes. This analysis shows clearly that
while there is frequently an essentially unique vacuum state for a 
static spacetime, in the dynamic case there is necessarily a one-parameter
family of vacuum solutions of the Schr\"odinger equation.

In sections 4 and 5, these techniques are illustrated using two simple
two-dimensional `cosmological models' which have been much
discussed in the literature. The first represents an asymptotically static 
universe that undergoes a period of expansion while the second
describes a universe which expands from an initial singularity before
recollapsing. These are of particular interest as models illustrating
cosmological particle creation. We solve the Schr\"odinger wave functional 
equation exactly to determine the candidate vacuum states and discuss
carefully the interpretation that particles are created by the expansion.

A number of technical results on solutions and Green functions for
various differential equations occurring in the text are given in the 
appendices.


\section{Schr\"odinger Picture in Minkowski Spacetime}

To introduce the Schr\"odinger formalism, we begin by reviewing the
wave functional equation and its solutions in Minkowski spacetime.
(For a review see, e.g. \cite{jackiw89}.)
Consider a massive scalar field \(\vphi(x)\) with action\footnote{Our 
conventions are as follows. For generality we
consider a spacetime of \(d+1\) dimensions. The metric
signature is \((+,-,\ldots,-)\). Coordinates for the spacetime point
\(x\) are denoted either by \((x^0,x^i)\), where \(i = 1,\ldots,d\),
or by \((t,\ux)\).}
\be
S = \int d^{d+1}x~{\cal L}\label{eq:action}
\ee
where the Lagrangian density is
\be
{\cal L} = \frac{1}{2} \Bigl\{\eta^{\m\n}(\pl_\m
{\vphi})(\pl_\n {\vphi})-m^2{\vphi}^2 \Bigr\}
\ee
The canonical momentum conjugate to the field is defined by
\be
{\pi}(x)=\frac{\pl{\cal L}}{\pl(\pl_0 {\vphi})}=
\pl_0 {\vphi}(x)
\ee
The Hamiltonian is constructed by a Legendre transform of the Lagrangian
\begin{eqnarray}
H(\pi, \vphi) & = & \int d^d\ux~[\pi\dot{\vphi}-{\cal L}]
\label{eq:legendre}\\
& = & \frac{1}{2}\int d^d\ux~\Bigl\{
{\pi}^2 - \eta^{ij} (\pl_i \vphi) (\pl_j \vphi)
+m^2{\vphi}^2 \Bigr\}
\end{eqnarray}
 
The system can now be canonically quantised by treating the fields as
operators and imposing appropriate commutation relations. This involves
the choice of a foliation of the spacetime into a succession of
spacelike hypersurfaces. In Minkowski spacetime, it is natural
to choose these to be the hypersurfaces of fixed \(t\) and impose
the equal-time commutation relations
\[
[\vphi(\ux,t), \pi(\uy,t)] =
i \delta^d(\ux - \uy)
\]
\be
[\vphi(\ux,t),\vphi(\uy,t)] =
[\pi(\ux,t),\pi(\uy,t)] = 0
\label{eq:etcommrel}
\ee
 
In the Schr\"odinger picture, we take the basis vectors of the
state vector space to be the eigenstates of
the field operator \(\vphi(x)\) on a fixed \(t\) hypersurface,
with eigenvalues \(\phi(\ux)\),
\be
\vphi(x)|\phi(\ux),t\rangle = \phi(\ux)|\phi(\ux),t\rangle
\label{eq:representation}
\ee
Notice that the set of field eigenvalues \(\phi(\ux)\) is independent of
the value of \(t\) labelling the hypersurfaces. This will be especially
important in the curved spacetime context.
 
In this picture, the quantum states are explicit functions of time
and are represented by wave functionals \(\Psi[\phi(\ux),t]\).
Operators \(O(\pi, \vphi)\) acting on these states may be represented
by
\be
O\Bigl(\pi(x), \vphi(x)\Bigr) ~  \sim ~  O\Bigl(-i\frac{\delta}
{\delta {\phi}(\ux)}, \phi(\ux)\Bigr)
\ee
consistent with the commutation relations (\ref{eq:etcommrel}).
The Schr\"odinger equation, which governs the evolution of the wave
functional between the spacelike hypersurfaces, is
\begin{eqnarray}
i{\frac {\pl\Psi[\phi,t]}{\pl t}}& = & H\Bigl
(-i\frac{\delta}{\delta \phi(\ux)},
\phi(\ux)\Bigr)\Psi[\phi,t] \nonumber \\
& = &\frac{1}{2} \int{d^d\ux}~
\Bigl\{ -\frac{\delta^2}{\delta \phi^2} ~-~
\eta^{ij}(\pl_i\phi) (\pl_j\phi)
~+~ m^2\phi^2\Bigr\} \Psi[\phi,t] \label{eq:swfemink}
\end{eqnarray}
 
\vspace{0.5cm}
To solve this equation, we make the ansatz that,
up to a time-dependent phase, the vacuum wave
functional is a simple Gaussian. This is in direct analogy with the
standard solution of the harmonic oscillator
in quantum mechanics. We therefore write
\be
\Psi_0[\phi,t] = N_0(t) \
\exp\left\{-\frac{1}{2}\int{d^d\ux}\int{d^d\uy} \
\phi(\ux)\,G(\ux,\uy)\,\phi(\uy)\right\}
 \,=\, N_0(t)\psi_0[\phi]
\label{eq:psi0mink}
\ee
Substituting into (\ref{eq:swfemink}) and comparing coefficients of
\([\phi^0]\) determines \(N_0(t)\) from
\be
\frac{d\ln N_0(t)}{dt} = -~ \frac{i}{2}
\int d^d\ux ~G(\ux,\ux)
\label{eq:timemink}
\ee
while comparing coefficients of \([\phi^2]\) gives an equation
for the kernel \(G(\ux,\uy)\):
\be
\int d^d\uz ~G(\ux,\uz)G(\uz,\uy) =
(- \nabla^2 + m^2)\delta^d(\ux - \uy)
\label{eq:kernelmink}
\ee
The kernel equation can be solved by using several methods, either by
working in momentum space, or coordinate space, or using the
Schwinger-DeWitt representation of the propagator 
\cite{schwinger51,dewitt65}.

We show here the momentum space approach, which is the most convenient
in Minkowski spacetime.
Introducing the Fourier transform of the kernel,
\be
  G(\ux, \uy) = \int \frac{d^d\uk}{(2\pi)^d}
  ~ e^{i\uk.(\ux-\uy)} \: \widetilde{G}(\uk)
\ee
the kernel eq.(\ref{eq:kernelmink}) reduces to
\be
  \widetilde{G}(\uk)^2 = \uk^2 + m^2
\ee
The inverse of the kernel, \(\Delta(\ux,\uy)\), is defined as
\be
  \int d^d\uz ~G(\ux,\uz)\Delta(\uz ,\uy)=
  \delta^d(\ux - \uy)
\ee
and is therefore given explicitly by
\be
  \Delta (\ux,\uy) = \int
  \frac{d^d\uk}{(2\pi)^d}
  ~ \frac{e^{i\uk.(\ux-\uy)}}
  {\sqrt{\uk^2 + m^2}}
\ee

The kernel is simply related to the Green functions of the theory
(see appendix B). A general Green function can be written as 
\be
{\cal G}(x,y) = \int \frac{d^{d+1}k}{(2\pi)^{d+1}}
  ~ \frac{e^{i\uk.(\ux-\ux')~ - ~ik_0(x_0 - y_0)}  } {k_0^2 -\w^2}
\ee
where \(\w^2={\uk^2 + m^2}\). The \(k_0\) integral has poles
at \(k_0=\pm\,\w\) and can be evaluated with different
boundary conditions, determining how the contour goes round the poles.
In particular, the Wightman function ${\cal G}_+(x,y)$ is
defined by choosing the contour integral to close around 
the \(k_0=\w\) pole as in Fig(\ref{fig:xfigwightman})
of appendix~\ref{sec-green}.
For equal times, this Green function reduces to
\begin{eqnarray}
  {\cal G}_+(x,y)|_{ET}&=&\frac{1}{2} ~ \int
  \frac{d^d\uk}{(2\pi)^{d}}
  ~ \frac{e^{i\uk.(\ux-\uy)} }
  {\sqrt{\uk^2 + m^2}}  \\
  &=& {1\over2\pi}\left({\frac{m}{2\pi|\ux-\uy|}}\right)^{{d\over2}
  -{1\over2}}
  K_{{d\over2}-{1\over2}}(m|\ux-\uy|)
\label{eq:invkernel}
\end{eqnarray}
The inverse kernel is therefore simply
twice the Wightman function evaluated at equal times, 
i.e.
\be
\Delta(\ux,\uy) = {2} {\cal G}_+(x,y)|_{ET}
\label{eq:invwightmanmink}
\ee
The reason for this identification is made clear in appendix~\ref{sec-vev}
where we discuss the representation of vacuum expectation values and two-point functions in the 
Schr\"odinger picture.
 
The kernel itself is
\begin{eqnarray}
  G(\ux,\uy) &=&\frac{1}{2} {\cal G}_+(x,y)|_{ET}^{-1} \\
  \nonumber \\
  \label{eq:kernelsolutionmink}
  &=&\int \frac{d^d\uk}{(2\pi)^d}\:\sqrt{\uk^2 + m^2}
  ~ {e^{i\uk.(\ux-\uy)}}  \\
  &=& -2 \left({\frac{m}{2\pi|\ux-\uy|}}\right)^{{d\over2}
  +{1\over2}}
  K_{{d\over2}+{1\over2}}(m|\ux-\uy|)
\end{eqnarray}
The time dependent phase factor in the vacuum wave functional
is
\be
  N_0(t) = e^{-iE_0t}
\ee
where
\begin{eqnarray}
  E_0 &=& \frac{1}{2}\int d^d\ux~
  G(\ux,\ux) \\
  &=&\frac{1}{2}\int d^d\uk ~\sqrt{\uk^2+m^2} ~~\delta^d(0)
\end{eqnarray}
is the (divergent) ground state energy.
 
\vspace{0.5cm}
 
There are several ways to derive the excited state solutions to
the Schr\"odinger wave functional equation, but with a view to
the curved spacetime generalisation, we start with a method
that does not presuppose a particle interpretation of the
excited states.
 
Making the following ansatz for the first excited state
\be
\Psi_1[\phi,t]=2 \,N_1(t) \int d^d\ux \int d^d\uy \:
  \phi(\ux)G(\ux,\uy)f(\uy)\:\psi_0[\phi]
~=~N_1(t) \psi_1[\phi]
\label{eq:excstate}
\ee
the Schr\"odinger equation is solved with the same kernel
(\ref{eq:kernelmink}) as for the ground state, provided
\be
  N_1(t) = e^{-iE_1t}
\ee
where the energy gap \(\Delta E_1 = E_1 - E_0\) satisfies
\be
  \Delta E_1 \int d^d\uy~G(\ux,\uy)f(\uy)
  =(-\nabla^2+m^2)f(\ux)
\ee
This is readily solved if  \(f(\ux)\) is an eigenfunction of the operator
\((-\nabla^2+m^2)\). We therefore choose
\(f_{\uk}(\ux)=e^{i\uk . \ux}\), which
has an eigenvalue of \((\uk^2+m^2)\). We then find
\[
\Delta E_1 \int d^d\uy ~G(\ux,\uy)
f_{\uk}(\uy)   ~=~ (\uk^2+m^2) f_{\uk}(\ux)
\]
which reduces to
\[
\Delta E_1 ~ \widetilde{G}(\uk) ~=~ \uk^2 + m^2
\]
so the energy gap is
\be
  \Delta E_1 =
  \sqrt{\uk^2 + m^2} = \w
\ee
Clearly this admits a particle interpretation where the first
excited state \(\Psi_1\) is a one-particle state
and \(E_1\) is its energy. This is apparent from
the energy gap, \(\Delta E_1 = \sqrt{\uk^2 + m^2}\),
which is the energy of one particle with momentum \(\uk\).
The wave functional carries a label \(\uk\), introduced through
the eigenfunction \(f_{\uk}\), which specifies the momentum of
the particle.

The Schr\"odinger equation can also be solved directly
(as in the quantum mechanical harmonic oscillator)
for an $n$-particle state, giving
\be
  \Psi_n[\phi, t] = e^{-iE_nt}H_n[\phi]
  \psi_0[\phi]
\ee
where the \(H_n\) are generalised Hermite polynomials defined
using the kernel and eigenfunctions \(f_{\uk_j}\) contracted with
the field \(\phi\) as in (\ref{eq:excstate}).
The energy of the \(n\)th excited state is simply
\be
  E_n  = E_0 +  \sum_{i=1}^{n} \w(\uk_i)
\ee
where \(\uk_i\) is the momentum of the \(i\)th particle.
 
The excited states can also be described using creation
and annihilation operators. We define
\begin{eqnarray}
  \hat{a}(\uk)
  &=&\int d^d\ux~
  \left[\int d^d\uy ~G(\ux,\uy) \phi(\uy)  + \frac{\delta}{\delta
    \phi(\ux)}\right] f_{\uk}^*(\ux)      \\
  &=&\int d^d\ux~
  \left[\w \phi(\ux) + \frac{\delta}{\delta
      \phi(\ux)}\right] f_{\uk}^*(\ux)
  \label{eq:annihilationmink}
\end{eqnarray}
and
\begin{eqnarray}
  \hat{a}^{\dagger}(\uk)
  &=&\int d^d\ux~
  \left[\int d^d\uy ~G(\ux,\uy) \phi(\uy) - \frac{\delta}{\delta
    \phi(\ux)}\right] f_{\uk}(\ux)      \\
  &=&\int d^d\ux~
  \left[\w \phi(\ux) - \frac{\delta}{\delta
      \phi(\ux)}\right] f_{\uk}(\ux)
  \label{eq:creationmink}
\end{eqnarray}
consistent with the commutation relations
\be
[\hat{a}(\uk)~ ,~ \hat{a}^{\dagger}(\underline{p})] =
2\,(2\pi)^d~ \w ~  \delta^d(\uk-\underline{p})
\label{eq:commrelmink}
\ee
These can be used iteratively to construct all the excited states,
starting from
\be
  \hat{a}(\uk)\psi_0 = 0 ~~~~~~~~~~~~~
  \hat{a}^{\dagger}(\uk)\psi_0=\psi_1
\label{eq:creation}
\ee
The $n$th excited state is labelled by the $n$ particle momenta, 
viz.
\begin{equation}
  \psi_n(\uk_1,\dots,\uk_n) =  
  a^\dagger(\uk_1) \dots  a^\dagger(\uk_n) \psi_0
\end{equation}

Orthogonality relations between these states follow directly from
the commutation relations of the creation and annihilation operators.
We have
\begin{eqnarray}
   \langle \psi_n(\uk_1,\dots ,\uk_n)| \psi_m(\up_1,\dots,\up_m)  \rangle
   &=& \delta_{mn}\Big\{\prod_{i=1}^n 2 \w_i (2\pi)^d \delta^d(\uk_i - \up_i)
     \nonumber  \\
   &&  + \dots n! \ \ \mbox{perms} \Big\}\langle \psi_0 |\psi_0 \rangle
\label{eq:phinphimmink}
\end{eqnarray}
where the inner product is defined by
\be
\langle \Psi_1 | \Psi_2 \rangle 
~=~ \int {\cal D}\phi~ \Psi_1^*(\phi) \Psi_2(\phi)
\ee
and $\w_i = \sqrt{\uk_i^2 + m^2}$.
The Hamiltonian and particle number operators are written in terms
of $a$ and $a^{\dagger}$ as usual as
\begin{equation}
  \label{eq:anncrhamiltonianmink}
  H = \frac{1}{4} \int \frac{d^d\uk}{(2\pi)^d} 
  \left[\hat{a}^{\dagger}(\underline{k})\hat{a}(\underline{k})+
    \hat{a}(\underline{k})\hat{a}^{\dagger}(\underline{k})\right]
\end{equation}
and
\begin{equation}
  N = \int  \frac{d^d\uk}{(2\pi)^d}  \frac{1}{2\w}~
    \hat{a}^{\dagger}(\underline{k})\hat{a}(\underline{k})
\label{eq:numberopmink}
\end{equation}

It will often be convenient to express the wave functional for the
vacuum and excited states in terms of the Fourier transform
$\tphi(\uk)$ of the field eigenvalues defined in eq.(\ref{eq:representation}).
We define
\begin{equation}
  \label{eq:fieldmodesmink}
  \phi(\ux) = \int \frac{d^d\uk}{(2\pi)^d}\: e^{i\uk.\ux} 
  \ \tphi(\uk)
\end{equation}
and since $\phi(\ux)$ is real, ${\tphi}^\ast(\uk) = \tphi(-\uk)$. 
The vacuum wave functional is therefore
\begin{equation}
  \label{eq:wavefnalmomentummink}
  \psi_0[\tphi] = 
  \exp\left\{-\half  \int \frac{d^d\uk}{(2\pi)^d} \:\w\: \: |\tphi(\uk)|^2\right\}
\end{equation}
In terms of $\tphi(\uk)$, the creation and annihilation operators 
simplify: 
\begin{eqnarray}
  \hat{a}(\uk) &=& \left[\w \: \tphi(\uk) + 
    \frac{\delta}{\delta {\tphi}^\ast(\uk)}\right] \\
  \hat{a}^\dagger(\uk) &=& \left[\w \: {\tphi}^\ast(\uk) -
    \frac{\delta}{\delta \tphi(\uk)}\right]
\end{eqnarray}
where ${\delta \tphi(\up)}/{\delta \tphi(\uk)} = 
(2\pi)^d \delta^d(\uk-\up)$ and 
${\delta \tphi^*(\up)}/{\delta \tphi(\uk)} = 
(2\pi)^d \delta^d(\uk+\up)$ .
The excited states for particles with momenta ${\uk}_i$ are therefore given by
\begin{eqnarray}
  \label{eq:wavefnalexcitedmomentummink}
  \psi_1 &=& 2 [\w_1 {\tphi}^\ast(\uk_1)] \psi_0 \\
  &&\nonumber \\
  \psi_2 &=& \left\{4 [ \w_1 {\tphi}^\ast(\uk_1)]  [\w_2 {\tphi}^\ast(\uk_2)] 
    -2 \w_1 (2\pi)^d \delta^d(\uk_1+\uk_2)\right\} \psi_0 
\end{eqnarray}
etc.
The general $n$-particle wave functional is expressed in terms of 
generalised Hermite polynomials with $n$ momentum labels, viz.
\begin{eqnarray}
  \label{eq:hermitemink}
   \Psi_n[\widetilde\phi,t;\uk_1, \dots,\uk_n] &=& e^{-iE_nt} 
   H_n[\tphi;\uk_1, \dots,\uk_n]  \psi_0[\tphi]\nonumber \\
   && \nonumber \\
   &=& e^{-iE_nt} \sum_{m=0}^{\left[\sfrac{n}{2}\right]}
   \Big\{(-1)^m 2^{n-m} \left[\prod_{i=1}^{m}\w_{2i-1} (2\pi)^d 
     \delta(\uk_{2i-1} + \uk_{2i})\right]\nonumber \\
   && \left[\prod_{j=2m+1}^{n} \w_j {\tphi}^\ast(\uk_j)
   \right] + \dots\mbox{$(2m-1)!! {n\choose2m}$}
   \mbox{perms}\Big\}\psi_0
\end{eqnarray}
where there are $(2m-1)!! {n\choose2m}$ permutations of the pairing of 
momenta at each stage of the sum. The sum is from zero to 
$\left[\sfrac{n}{2}\right]$-
which is the largest integer $\leq \frac{n}{2}$. 
Products $\prod_{i=a}^b$ with $b<a$ are defined to be 1.

\vspace{0.3cm}
All this analysis is based on the specific foliation of Minkowski
spacetime in which the family of spacelike hypersurfaces \(\S\)
are equal-time hypersurfaces and the Hamiltonian evolution is
along the timelike Killing vector field \(\pl \over \pl t\).
While this is convenient, it is by no means a unique choice.
In section 3, we set up the Schr\"odinger equation
for an arbitrary foliation in curved spacetime.
This formalism may of course be specialised to Minkowski spacetime
by simply replacing the metric \(g_{\m\n}\) by \(\eta_{\m\n}\)
throughout.


\subsection{General Gaussian state}
\label{sec-gaussianmink}

In the discussion of the cosmological models in sections 4 and 5 we encounter
Gaussian states where the kernel is time dependent and not simply
equal to \(\w\), the eigenvalue of the wave operator. In anticipation of this,
we show here how such a state in Minkowski spacetime may be expressed
as a superposition of excited many-particle states.

Consider a solution of the Schr\"odinger equation of the form
\be
  \Psi[\tphi,t] = N(t)
  \exp\left\{-\half  \int \frac{d^d\uk}{(2\pi)^d} 
    \:\Omega(\uk;t)\: \: |\tphi(\uk)|^2\right\}
\ee
where 
\be
N(t) = \exp \left\{-{i\over2}\int^t dt \int {d^d\uk\over (2\p)^d}\Omega(\uk;t)\right\}
\ee
and \(\Omega(\uk;t)\) is a solution of
\be
  i\frac{\partial \Omega(\uk;t)}{\partial t} = \w^2 - \Omega^2
\ee
This can be written as a linear superposition of \(n\)-particle states
as follows:
\be
  \Psi = \sum_{n=0}^\infty 
  \left[\prod_{i=1}^{n} \int  \frac{d^d\uk_i}{(2\pi)^d}\right]
  c_n(\uk_1,\dots,\uk_n) \psi_n(\uk_1,\dots,\uk_n)
  \label{eq:expansionmink}
\ee
where the (time-dependent) expansion coefficients are
\be
  c_n(\uk_1,\dots,\uk_n) = 
  \frac{\int {\cal D}\tphi \psi_n^\ast(\uk_1,\dots,\uk_n)\Psi}
  {n!\left[\prod_{i=1}^{n} 2 \w_i\right]}
\ee
and we are assuming constant normalisation factors are included such
that \(\langle\psi_0|\psi_0\rangle\) and \(\langle\Psi|\Psi\rangle\) are both 1.
(For the justification that this is possible for the general Gaussian state,
see section 3.)
The \(n\)-particle states have been given earlier. We therefore have
\be
\int {\cal D}\tphi \psi_n^\ast(\uk_1,\dots,\uk_n)\Psi = 
\int {\cal D}\tphi \: H_n[\tphi;\uk_1, \dots,\uk_n]  \:
e^{-{1\over2}\int \frac{d^d\uk}{(2\pi)^d} \:(\Omega + \w)\: \: |\tphi(\uk)|^2}
\ee
The odd moments of the integral vanish and therefore the coefficients 
$c_n$ vanish for $n$ odd. For even $n$, we find
\begin{eqnarray}
  c_n(\uk_1,\dots,\uk_n) &=& \Big\{\frac{1}{n!}
    \left[\prod_{i=1}^{{n}/{2}} \frac{(2\pi)^d 
        \delta^d(\uk_{2i-1} + \uk_{2i})}{2 \w_{2i-1}}
      \left(\frac{\w_{2i-1}-\Omega_{2i-1}}{\w_{2i-1}+\Omega_{2i-1}}
      \right)\right]\nonumber \\
    & & \ \ + \dots (n-1)!! \ \mbox{perms} \dots \Big\}
   \langle \psi_0|\Psi \rangle
\end{eqnarray}
where $\Omega_i = \Omega(\uk_i;t)$ and the $(n-1)!!$ permutations 
specify different combinations of the paired momenta. 

It is particularly interesting to look at the expectation value in the general
Gaussian state of the number operator \(N(\uk)\), which counts the number 
of particles with prescribed momentum in the 
expansion (\ref{eq:expansionmink}). This is
\begin{eqnarray}
 \langle \Psi|N(\uk)|\Psi \rangle
  &=&    \frac{1}{2\w} 
  \langle \Psi|\hat{a}^{\dagger}(\underline{k})\hat{a}(\underline{k})|
    \Psi \rangle  \nonumber \\
  &=& \half \left\{ 
    \w \langle \Psi|\tphi(\uk)\tphi(-\uk)|\Psi \rangle - 
    {1\over\w}\langle \Psi|\frac{\delta^2}
      {\delta\tphi(\uk) \delta\tphi(-\uk)}|\Psi \rangle -
    (2\pi)^d \delta^d(0)\right\}  \nonumber \\
   &{}&{}
\end{eqnarray}
The expectation values can be easily evaluated using techniques described
in appendix C. For complex \(\Omega\), we find the expectation value
of the number density to be
\be
    \langle {\cal N}(\uk) \rangle =
     \frac{(\w-\Omega)(\w-{\Omega}^*)}{2\w(\Omega+{\Omega}^*)}
\ee
This result can also be recovered using the explicit expressions for
the expansion coefficients \(c_n\) given above.

\newpage

\section{Schr\"odinger Picture in Curved Spacetime}

We consider globally hyperbolic spacetimes of topology
\({\cal M} = {\cal R} \times \S\) with a global timelike
vector field. To provide a general foliation of the
spacetime, we use the formalism of embedding variables 
\cite{kuchar76,freese85,halliwell91}.
We therefore choose a family of spacelike hypersurfaces \(\S\),
with intrinsic coordinates \(\xi^i\),
labelled by a `time' parameter \(s\) and consider evolution
along the integral curves of \(\pl\over\pl s\).
The embeddings characterising the foliation are 
maps \(x:\S\rightarrow{\cal M}\) which take a point on the
hypersurface \(\S\) to a spacetime point \(x^\m = x^\m(s,\xi^i)\).
 
First we need to construct the projections of spacetime quantities
normal and tangential to the hypersurface \(\S\). The tangential
projections are defined using
\be
  p^{\m}_i = \frac{\pl x^{\m}}{\pl \xi^{i}}
\label{eq:ev1}
\ee
The normal to the surface,  \(n_{\m}\), is timelike and
is uniquely defined by
\be
  n_{\m} p^{\m}_i = 0
\label{eq:ev2}
\ee
with \(g^{\m \n} n_\m n_\n = 1\).
Any spacetime tensor can be expanded in terms of its projections
normal and tangential to \(\S\).
We can also introduce an induced metric on \(\S\),
\be
  h_{ij}=g_{\m \n} p^{\m}_{i} p^{\n}_{j}
\label{eq:ev3}
\ee
and raise and lower hypersurface indices, e.g.
define \(p_\m^i = h^{ij} p_j^\n g_{\m\n} \).
 
Now consider the deformation vector of the foliation, defined
as
\begin{eqnarray}
  N^{\m} &=& \dot{x}^{\m} \equiv \frac{\pl x^{\m}}{\pl s}
  \nonumber \\
  &=& N n^{\m} + N^i p^{\m}_{i}
  \label{eq:ev4}
\end{eqnarray}
where \(N\) and \(N^i\) are the `lapse' and `shift' functions of the
embeddings given by
\begin{eqnarray}
  N & = & n_{\m} N^\m
  \label{eq:ev5}\\
  N^i & = & p_{\m}^{i} N^\m
  \label{eq:ev6}
\end{eqnarray}
In terms of these functions, the metric interval can be written as
\be
  g_{\m\n}dx^{\m}dx^{\n}
  ~=~ (N^2 + N^i N_i)ds^2
  + 2N_i ds d\xi^i + h_{ij} d\xi^i d\xi^j
\label{eq:ev7}
\ee
We also need the inverse relations
\begin{eqnarray}
  \frac{\pl s}{\pl x^{\m}} &=& \frac{n_{\m}}{N}
  \label{eq:ev8}\\
  \frac{\pl \xi^i}{\pl x^{\m}} &=& \frac{-n_{\m}N^i}{N}+p^i_{\m}
\label{eq:ev9}
\end{eqnarray}
and the Jacobian
\be
  {\rm det}   \Bigl[\frac{\pl x^{\m}}{\pl(s,\xi^i)}\Bigr] =
  \frac{N\sqrt{-h}} {\sqrt{-g}}
\label{eq:ev10}
\ee
 
\vspace{0.3cm}
With these preliminaries complete,
we can now construct the Hamiltonian describing the evolution
appropriate to this foliation, i.e.~along the integral curves
of the timelike vector field \({\pl\over\pl s}\).
We consider a massive scalar field theory coupled to gravity.
The action is
\be
  S = \half \int_{\cal M} ~  d^{\,d+1}x~\sqrt{-g} \Bigl[ g^{\m\n}
  \frac{\pl \varphi}{\pl x^{\m}}\frac{\pl \varphi}{\pl
    x^{\n}} -(m^2+\xi R)\varphi^2\Bigr]
\label{eq:ev11}
\ee
and so in terms of the embedding variables,
\begin{eqnarray}
  S &=& \half \int_{R}ds  ~ \int_{\S} d^d\underline{\xi} ~
  {\rm det} \Bigl[\frac{\pl
    x^{\m}}{\pl(s,\xi^i)}\Bigr]\sqrt{-g}\Bigl\{ g^{\m\n}\Bigl[
  \frac{\pl s}{\pl x^{\m}}\frac{\pl s}{\pl
    x^{\n}}\dot{\varphi}^2 \nonumber\\ &&\nonumber\\ &&+2\frac{\pl
    s}{\pl x^{\m}}\frac{\pl \xi^i}{\pl x^{\n}}
  \dot{\varphi}(\pl_i\varphi) +\frac{\pl \xi^i}{\pl
    x^{\m}}\frac{\pl \xi^j}{\pl x^{\n}}
  (\pl_i\varphi)(\pl_j\varphi)\Bigl]-(m^2+\xi R)\varphi^2\Bigr\}
\label{eq:ev12}
\end{eqnarray}
where \(\dot\varphi \equiv {\pl\varphi\over\pl s}\) and
\(\pl_i\varphi \equiv {\pl\varphi\over \pl\xi^i}\).
Using eqs.(\ref{eq:ev8}), (\ref{eq:ev9}) and
(\ref{eq:ev10}), the action can be written in the form
\be
  S = \int_{R}ds ~  \int_{\S}d^d\underline{\xi} ~~    {\cal L}
\ee
where the Lagrangian density is,
\be
  {\cal L} = \half N \sqrt{-h}\Bigl\{\frac{\dot{\varphi}^2}{N^2} -
  \frac{2N^i}{N^2}\dot{\varphi}(\pl_i\varphi) +
  \Bigl[\frac{N^iN^j}{N^2}+h^{ij}\Bigr](\pl_i\varphi) (\pl_j\varphi)
  -(m^2+\xi R)\varphi^2\Bigr\}
\ee
The conjugate momentum is defined as usual to be
\be
  \pi = \frac{\pl{\cal L}}{\pl \dot{\varphi}} =
  \frac{\sqrt{-h}}{N}\Bigl\{\dot{\varphi}-N^i(\pl_i\varphi)\Bigr\}
\label{eq:conjphi}
\ee
The Hamiltonian is given by the Legendre transform
\begin{eqnarray}
  H & = & \int_{\S} d^d\underline{\xi}~[  \pi\dot{\varphi}-{\cal L}]\\
  & & \nonumber \\
  & = & \int_{\S} d^d\underline{\xi}~[  N{\cal H}+N^i{\cal H}_i]
\end{eqnarray}
where
\begin{eqnarray}
  {\cal H} &=& \half \sqrt{-h}\Bigl\{-\frac{\pi^2}{h} -
  h^{ij}(\pl_i\varphi) (\pl_j\varphi) + (m^2+\xi R)\varphi^2\Bigr\}\\
  {\cal H}_i &=& \pi (\pl_i\varphi)
\end{eqnarray}
 
Canonical quantisation is implemented by imposing
equal-\(s\) commutation relations for the field operators on the
hypersurface \(\S\), viz.
\[
[\varphi(\underline{\xi},s), \pi(\underline{\zeta},s)] = i \delta^d
(\underline{\xi} - \underline{\zeta})
\]
\be
  [\varphi(\underline{\xi},s),\varphi(\underline{\zeta},s)] =
  [\pi(\underline{\xi},s),\pi(\underline{\zeta},s)] = 0
\ee
In the Schr\"odinger picture, the quantum states are represented by
wave functionals \(\Psi[\phi(\underline{\xi}), s; N, N^i, h_{ij}]\)
where the variables \(\phi(\underline{\xi})\) are the eigenvalues of the
field operator on the equal-\(s\) hypersurfaces, i.e.
\be
\varphi(x)|\phi(\underline{\xi}),s\rangle =
\phi(\underline{\xi})|\phi(\underline{\xi}),s\rangle
\ee
As before, the commutation relations are realised by representing
operators in terms of the field eigenvalues \(\phi(\underline{\xi})\)
and functional derivatives \(\d\over\d\phi\), viz.\footnote{The 
functional derivative is defined so that
\(\d\phi(\uy)/\d\phi(\ux) = \d^d(\ux - \uy)\).
The delta function density is given by
\( \d^d(\ux,\uy) = (\sqrt{-h_x})^{-1} \d^d(\ux - \uy) \)
and satisfies \(\int d^d\ux \sqrt{-h_x} \d^d(\ux,\uy)
f(\ux) = f(\uy)\).}
\be
   O\Bigl(\pi(x), \varphi(x)\Bigr) \sim
   O\Bigl(-i{\d\over\d\phi(\ux)}, \phi(\ux) \Bigr)
\ee
The Schr\"odinger wave functional equation gives the evolution
of \(\Psi[\phi(\underline{\xi}),s]\)~:
\be
  i{\frac {\pl\Psi}{\pl s}}= ~ \int_{\S}
  {d^d\underline{\xi}}~\Bigl\{ \half N \sqrt{-h} \Bigl( \frac{1}{h}
  \frac{\delta ^2}{\delta \phi ^2} -h^{ij}\pl_{i}
  \phi\pl_j\phi+(m^2+\xi R)\phi^2 \Bigr) - i N^i \pl_i\phi
  \frac{\delta}{\delta \phi}\Bigr\} \Psi
\label{eq:swfemanifest}
\ee
 
\vspace{0.3cm}
This equation holds for an arbitrary foliation
(specified by \(N\), \(N^i\) and \(h_{ij}\))
and makes explicit the foliation dependence of the wave functional.
In practical applications, however, it is often convenient to choose
spacetime coordinates which reflect the desired foliation.
In this case, we simply identify the embedding variables
\((s, \underline{\xi})\) with the spacetime coordinates \((t, \ux)\).
The lapse and shift functions are then \(N= \sqrt{g_{00}}\)
and \(N^i = 0\) (so that \(g_{0i}=0\))
while the induced metric is just
\(h_{ij} = g_{ij}\). The Schr\"odinger equation reduces to
\be
  i{\frac {\pl\Psi}{\pl t}}~=~  \half \int{d^d\ux}~
  \sqrt{-g} ~ \Bigl\{ {g_{00}\over g}
  \frac{\delta ^2}{\delta \phi ^2}  -   g^{ij}(\pl_{i}
  \phi)(\pl_{j}\phi)+(m^2+ \xi R)\phi^2  \Bigr\} \Psi
  \label{eq:swfecst}
\ee
This is the form we shall usually use in the remainder of
this paper.
 
\vspace{0.3cm}
We can search for solutions to this Schr\"odinger equation
in analogy to those found in Minkowski space.
For the vacuum state, we make the ansatz
\be
  \Psi_0[\phi(\ux),t]=N_0(t)\psi_0[\phi,t]
\label{eq:wavecst}
\ee
where
\be
  \psi_0[\phi,t]=  \exp\left\{{-{1\over2}   \int d^d\ux
    \sqrt{-h_x}  \int d^d\uy    \sqrt{-h_y} ~
    \phi(\ux)   G(\ux, \uy ;t) \phi(\uy)}\right\}
\label{eq:wave2cst}
\ee
This has the same form as before, except that the kernel may now
depend on the `time' \(t\) since the metric is in general time
dependent.
The spatial integrals include the correct measure to
ensure that they are invariant \(d\) dimensional volume elements.
 
Substituting this ansatz into the functional Schr\"odinger
equation (\ref{eq:swfecst}) results in the time dependence equation
\be
  \frac{d \ln N_0(t)}{dt} = - {i\over2} \int d^d\ux ~
  \sqrt{-h_x} \sqrt{g_{00}^x} ~ G(\ux,\ux;t)
\label{eq:timecst}
\ee
and the kernel equation
\begin{eqnarray}
  i\frac{\pl}{\pl t}\left( \sqrt{h_x h_y} ~
    G(\ux,\uy;t) \right) &=& ~
  \int{d^d\uz ~ \sqrt{-h_z}\sqrt{g_{00}^z}
   ~\sqrt{h_x h_y}}
  ~ G(\ux,\uz;t)G(\uz,\uy;t)
  \nonumber \\ && \nonumber \\ &&- ~
  \sqrt{h_x h_y} \sqrt{g_{00}^x}
   ~(\Box_i + m^2 + \xi R)_x ~
  \delta^d(\ux ,\uy)
\label{eq:kernelcst}
\end{eqnarray}
where $\Box_i = \frac{1}{\sqrt{-g}}\pl_i( g^{ij} \sqrt{-g}
\pl _j)$ is the spatial part of the Laplacian (see appendix A).
Notice that this assumes that on the boundaries of the hypersurface
\(\S\) the fields or their derivatives vanish, since the surface terms
in an integration by parts have been omitted.
 
\vspace{0.3cm}
We can also mimic the Minkowski solutions for excited states,
although in the general case with explicit \(t\) dependence
in all the quantities the physical interpretation is not
immediately evident. Nevertheless, we can look for solutions of
the form
\begin{eqnarray}
  \Psi_1[\phi,t]&=& N_1(t) \psi_1[\phi,t] \\
  &=& 2N_1(t) \int d^d\ux\sqrt{-h_x}
  \int d^d\uy\sqrt{-h_y} ~
  \phi(\ux)  G(\ux,\uy;t)  f_{(\l)}(\uy) ~
  \psi_0[\phi,t] \nonumber \\
   &{}&{} 
\label{eq:firstwavecst}
\end{eqnarray}
This solves the Schr\"odinger equation with the same kernel
as for the vacuum state. The analogue of the energy gap equation is
\be
  \Delta E_1(t)~\int d^d\uy \sqrt{-h_y} ~
  G(\ux,\uy;t)f_{(\l)}(\uy)
  = \sqrt{g_{00}^x} ~ (\Box_i + m^2 + \xi R)_x
  ~f_{(\l)}(\ux)
\label{eq:gapeqncst}
\ee
where
\be
  \Delta E_1(t) = i\frac{d}{dt}\Bigl[\ln \frac{N_1}{N_0} \Bigr]
\label{eq:gapcst}
\ee
The nature of these states depends on the functions
\(f_{(\l)}(\ux)\). These may be chosen to be
eigenfunctions of the operator \((\Box_i+m^2+\xi R)\),
the suffix \( \l \) denoting the set of quantum numbers specifying the
degenerate eigenfunction (the analogues of the momenta \(\uk\) in
Minkowski spacetime).
 
We can also write down analogues of the creation and annihilation
operators of section 2.
If we write
\be
  \hat{a}(\l;t)
   = \int d^d\ux    \sqrt{-h_x}
  \left[\int d^d\uy  \sqrt{-h_y}~ G(\ux,\uy;t) \phi(\uy)~
  +~\frac{1}{\sqrt{-h_x}} ~ \frac{\delta}{\delta
    \phi(\ux)}\right]f_{(\l)}^*(\ux)
\ee
and
\be
   \hat{a}^{\dagger}(\l;t)
   = \int d^d\ux    \sqrt{-h_x}
  \left[\int d^d\uy  \sqrt{-h_y}~ G(\ux,\uy;t) \phi(\uy)~
  -~\frac{1}{\sqrt{-h_x}} ~ \frac{\delta}{\delta
    \phi(\ux)}\right]f_{(\l)}(\ux)
\ee
then
\be
  \hat{a}(\l;t) \psi_0[\phi,t] = 0 ~~~~~~~~~~~~~
  \hat{a}^{\dagger}(\l;t) \psi_0[\phi,t] = \psi_1[\phi,t]
\ee
and
\be
[\hat a(\l;t), \hat a^{\dagger}(\r;t)] =
2 \int d^d\ux \sqrt{-h_x} \int d^d\uy \sqrt{-h_y}~
f_{(\l)}^*(\ux) G(\ux,\uy;t) f_{(\r)}(\uy)
\ee
However, in contrast to Minkowski spacetime,
there is no simple particle interpretation
associated with these operators for a general spacetime.
 
\vspace{0.3cm}
The kernel equation (\ref{eq:kernelcst}) is difficult to solve
in general, so we shall specialise to three cases of particular
interest -- `static' spacetimes, where the metric is a function
only of the space coordinates, `dynamic' or Bianchi type I spacetimes,
where the metric depends only on the time coordinate,
and a special class of conformally static metrics of sufficient
generality to include all Robertson-Walker spacetimes.
Before restricting to these special cases, however, it is of
some interest to give a partial solution valid in general.
 
This partial solution is given in terms of a solution
\(\psi(x)\) to the wave equation
\be
  \sqrt{-g}[\Box + m^2 + \xi R] ~  \psi(x)=0
\ee
Any \(G(\ux,\uy;t)\) satisfying the equation
\be
  \int d^d\uy \sqrt{-h_y}\,G(\ux,\uy;t)\psi(t,\uy)
  ~=~  -i\sqrt{g^{00}_x}\frac{\pl}{\pl t}\psi(t,\ux)
\label{eq:kernelsolutiongeneral}
\ee
is then a solution of the full kernel eq.(\ref{eq:kernelcst}).
 
To see this, multiply the kernel equation
by \(\psi(t,\uy)\) and integrate with
respect to \(\uy\) and then substitute (\ref{eq:kernelsolutiongeneral}).
This gives
\begin{eqnarray}
  i\int d^d\uy ~\psi(t,\uy)  \frac{\pl}{\pl t}\left[
    \sqrt{h_x h_y} ~G(\ux,\uy;t)\right]
  &+&i\int d^d\uy  \sqrt{h_x h_y} ~G(\ux,\uy;t)
    \frac{\pl}{\pl t}\left[\psi(t,\uy)\right]\nonumber \\
  &=&-\sqrt{-g_x}(\Box_i + m^2 + \xi R)_x\,\psi(t,\ux)\nonumber
\end{eqnarray}
and combining the two differential terms we find
\[
i\frac{\pl}{\pl t}\left[ \int d^d\uy  \sqrt{-h_x}  \sqrt{-h_y} ~
G(\ux,\uy;t)  \psi(t,\uy)\right]
+\sqrt{-g_x}(\Box_i + m^2 + \xi R)_x ~\psi(t,\ux)=0
\]
Using eq.(\ref{eq:kernelsolutiongeneral}) for the kernel again
now gives
\[
\frac{\pl}{\pl t}\left(
  g^{00}_x  \sqrt{-g_x}\frac{\pl}{\pl t}\psi(t,\ux)\right)
+\sqrt{-g_x}(\Box_i + m^2 + \xi R)_x ~\psi(t,\ux)=0
\]
which is just the wave equation satisfied by \(\psi(t,\ux)\).
 
Later, when we have found explicit solutions for static or dynamic
spacetimes, we see that these may be found from
eq.(\ref{eq:kernelsolutiongeneral})
by writing the function \(\psi(x)\) as an appropriate Fourier
transform, using eqs.(\ref{eq:fieldstatic}) or
(\ref{eq:fielddynamic}) respectively,
and using orthonormality and completeness relations where applicable.

\subsection{Static spacetimes}

For static spacetimes, the kernel equation is solved with
a time-independent kernel satisfying
\be
  \int d^d\uz   \sqrt{-g_z}
  ~ G(\ux, \uz) ~G(\uz,\uy)
  =\sqrt{g_{00}^x}(\Box_i + m^2 + \xi R)_x ~\delta^d(\ux, \uy)
\label{eq:kernelstatic}
\ee
 
The solutions can be expressed in terms of the eigenfunctions of
\((\Box_i + m^2 + \xi R)\), which
are the Fourier transformed solutions \(\tpsi_{(\l)}(\w,\ux)\)
of the wave equation for a static metric, viz.
\be
  (\Box_i +  m^2 +\xi R)\tpsi_{(\l)}(\w,\ux)=
g^{00}\w^2(\l) \tpsi_{(\l)}(\w,\ux)
\label{eq:eigenstatic}
\ee
The eigenfunctions are specified by a set of discrete or continuous
quantum numbers \( \l \)
which generalise the momentum label \(\uk\) in flat space.
The eigenvalue \(\w(\l)\) is a function of the \(\l\).
These eigenfunctions satisfy orthonormality and completeness
relations described in appendix A.
 
It is then straightforward to check that the kernel equation
is solved by
\be
  G(\ux,\uy)=\sqrt{g^{00}_x\,g^{00}_y} \int \frac{d\m(\l)}{(2\pi)^d} 
  ~\w(\l)~  \tpsi_{(\l)}(\w,\ux)\tpsistar_{(\l)}(\w,\uy)
\label{eq:kerstatic}
\ee
In passing, we also note that this expression follows
by expressing \(\psi(x)\) in eq.(\ref{eq:kernelsolutiongeneral})
as a Fourier transform and using the completeness relation
(\ref{eq:compstatic}). It is also clear that the kernel is a
real function.
 
In Minkowski spacetime, we showed that the kernel was related to the
inverse Wightman function evaluated at equal time. We can derive a
similar result for a general static metric.
Introducing the inverse of the kernel, \(\Delta(\ux,\uy)\), defined
by
\be
\int d^d\uz  \sqrt{-h_z} ~\Delta(\ux,\uz)\,G(\uz,\uy)=
      \delta^d(\ux,\uy)
\ee
we find using the orthonormality and completeness relations
that
\be
  \Delta(\ux,\uy)= \int \frac{d\m(\l)}{(2\pi)^d} ~
  \frac{1}{\w(\l)} ~\tpsi_{(\l)}(\w,\ux)\tpsistar_{(\l)}(\w,\uy)
\label{eq:invkerstatic}
\ee
The Wightman function is evaluated in appendix B, eq.(\ref{eq:wightman})
and as expected we confirm
\be
  \Delta(\ux,\uy) = 2\,{\cal G}_+(x,y)|_{ET}
\label{eq:invwightmanstatic}
\ee
Again the reason for this identification is given in
appendix~\ref{sec-vev}.

The complete Schr\"odinger vacuum wave functional can therefore
be written as
\be
\Psi_0[\phi,t]  =  N_0(t)\psi_0[\phi]
\ee
where
\be
\psi_0=\exp\left\{\!-\!\! \! \int\!\!\frac{d\m(\l)}{(2\pi)^d}
\frac{\w(\l)}{2}\!\!\int\! \!d^d\ux \mbox{$\sqrt{-h_x}$}\sqrt{g_x^{00}}
\phi(\ux)\tpsi_{(\l)}(\w,\ux)\!\!
\int\! \!d^d\uy \sqrt{-h_y}\sqrt{g_y^{00}}
\phi(\uy) \tpsistar_{(\l)}(\w,\uy)\right\}
\ee
Notice how the fields $\phi(\ux)$ effectively project out
the kernel functions \(\tpsi_{(\l)}(\w,\ux)\)
appropriate to the boundary conditions on \(\phi\)
on the hypersurface \(\S\). We shall discuss this point in more detail
in the examples which follow.
The time dependent phase factor is
\be
N_0(t) = e^{-iE_0 t}
\label{eq:timestatic}
\ee
where \(E_0\) is the divergent vacuum energy,
\begin{eqnarray}
E_0 &=& {1\over2} \int d^d\ux \sqrt{-h_x}\sqrt{g_{00}^x}
~G(\ux,\ux)\\
&=& {1\over2} \int d\m(\l)~\w(\l)~\d^d(0)
\end{eqnarray}
 
\vspace{0.3cm}
The analysis given above of the excited states for a general
spacetime may now be applied. For a static spacetime, all
the quantities are time independent, so the states generated by
the creation and annihilation operators are genuine stationary states
with respect to the chosen Hamiltonian evolution.
 
The wave functional for the first excited state is
\be
\Psi_1[\phi,t] =  2N_1(t) \int d^d\ux\sqrt{-h_x}
  \int d^d\uy\sqrt{-h_y} ~
  \phi(\ux)  G(\ux,\uy)    f_{(\l)}(\uy) ~
  \psi_0[\phi]
\ee
where \(G(\ux,\uy)\) is the static kernel, \(N_1(t) = e^{-i\w(\l) t} N_0(t)\)
and \(f_{(\l)}(\uy)\) is a particular eigenfunction of
\((\Box_i + m^2 + \xi R) \) specifying the state.
The corresponding creation and annihilation operators simplify to
\be
  \label{eq:annihilationstatic}
  \hat{a}(\l)  = \int d^d\ux \left[
    \sqrt{-h_x}  \sqrt{g^{00}_x}~\w(\l) \phi(\ux) +
    \frac{\delta}{\delta \phi(\ux)}
  \right] f_{(\l)}^*(\ux)
\ee
\be
  \label{eq:creationstatic}
  \hat{a}^{\dagger}(\l) = \int d^d\ux \left[
    \sqrt{-h_x}  \sqrt{g^{00}_x}~\w(\l) \phi(\ux) -
    \frac{\delta}{\delta \phi(\ux)}
  \right] f_{(\l)}(\ux)
\ee
and satisfy the commutation relations
\be
[\hat{a}(\l), \hat{a}^{\dagger}(\r)] =
2 \w(\l)~ (2\pi)^d ~ \delta^d(\l,\r)
\ee

\goodbreak

\vspace{0.3cm}
As an elementary example, it is straightforward to check 
that in Minkowski spacetime choosing 
\(\tpsi_{(\l)}(\w;\ux) = \exp i\uk.\ux \), where
\(\w(\uk) = \sqrt{\uk^2 + m^2}\) and \(d\m(\l) = d^d\uk\), 
reproduces precisely the results of section 2.

\subsection{Dynamic spacetimes}

By a `dynamic' spacetime, we mean one where the metric depends only
on the `time' coordinate \(t\). The general 
metric, after rescaling the
\(t\) coordinate, can therefore be written as
\be
ds^2 = dt^2 - a_{ij}^2(t) dx^i dx^j
\ee
and defines a Bianchi type I spacetime \cite{tod90}.

To simplify the kernel equation for the special case of a dynamic
spacetime,  we introduce the Fourier transform \(\widetilde G(\uk,t)\)
(see appendix A for definitions),
\be
G(\ux,\uy;t) = \int {d^d\uk\over(2\pi)^d}~ e^{i\uk.(\ux-\uy)}~
\widetilde G(\uk;t)
\label{eq:ftdynamic}
\ee
The kernel equation reduces to
\be
i{\pl\over \pl t} \left[h \widetilde G(\uk;t)\right]
= h \sqrt{-g} \widetilde G^2(\uk;t)
+ \sqrt{-g} \omega^2(\uk;t)
\label{eq:partialG}
\ee
where \(\omega^2(\uk;t) = -g^{ij}k_i k_j + m^2 + \xi R \).
 
The solution\footnote{The {\schr} formalism
in flat Robertson-Walker spacetime has been extensively 
studied in refs.\cite{guven89,eboli89}.} may be expressed 
in terms of the functions
\(\tpsi(t,\uk)\) introduced in appendix~\ref{sec-wave}
 which satisfy the Fourier transform of the wave equation, viz.
\be
(\Box_0 - g^{ij}k_ik_j + m^2 + \xi R) \tpsi(t,\uk) = 0
\label{eq:box0}
\ee
We find
\be
\widetilde G(\uk;t) = {-i\over \sqrt{-g}}       ~{\pl\over\pl t}
{\rm ln} \tpsi(t,\uk)
\label{eq:tildeGsoln}
\ee
and thus for the kernel itself,
\be
G(\ux,\uy;t) = -i \sqrt{g^{00}}
{1\over \sqrt{-h}}
\int {d^d\uk\over (2\pi)^d} e^{i \uk.(\ux-\uy)}~
{\pl\over\pl t} {\rm ln} \tpsi(t,\uk)
\label{eq:kerndynamic}
\ee
Notice that this is again consistent with the general
expression (\ref{eq:kernelsolutiongeneral}) above.
 
To show this \cite{guven89}, we convert the non-linear first-order
differential equation for \(\widetilde G(\uk;t)\) into
a linear second-order equation.
Define
\be
F(\uk;t) = \exp \{-i\int \!dt\, a h\widetilde G(\uk;t)\}
\ee
where \(a = {\sqrt{-g}\over -h}\) and let
\(\dot a = {\pl a\over \pl t}\).
We then have
\be
{\pl^2\over\pl t^2} F
= -i a {\pl\over\pl t}(h\widetilde G)
-a^2 h^2\widetilde G^2 F - i \dot a h \widetilde G F
\ee
and using eq.(\ref{eq:partialG}),
\be
{\pl^2\over \pl t^2}F =  {\dot a\over a} {\pl\over\pl t} F
-a \sqrt{-g} \omega^2 F
\ee
However, this is just the equation satisfied by the functions
\(\tpsi(t,\uk)\), since by expanding \(\Box_0\) in
eq.(\ref{eq:box0}) we see
\be
\left(\pl_0^2 - {\dot a\over a} \pl_0 + a \sqrt{-g} \omega^2 \right)
\tpsi(t,\uk) = 0
\ee
We therefore identify \(F(\uk;t)\) with \(\tpsi(t,\uk)\)
and eq.(\ref{eq:tildeGsoln}) follows.
 
\vspace{0.3cm}
The function \(\tpsi(t,\uk)\) is the solution of a second-order
ordinary differential equation and is therefore a linear
combination of two independent solutions with arbitrary
coefficients. Since the kernel itself is the derivative of the
logarithm of \(\tpsi(t,\uk)\), the overall normalisation is
unimportant. The kernel therefore depends on one arbitrary
parameter.
 
This is a physically important difference from the static spacetime,
where the kernel is essentially unique. For a dynamic spacetime,
the `vacuum' states described by the Gaussian solution of the
Schr\"odinger wave functional equation form a one-parameter
family \cite{guven89}. The selection of one of these as the vacuum
must then be made on further physical grounds. We shall return
to this point in the context of particular examples later.

The Schr\"odinger vacuum wave functional can therefore be written
as
\be
\Psi_0[\phi,t] = N_0(t) \psi_0[\phi,t]
\ee
where
\be
\psi_0 = \exp \left\{{i\over2} \sqrt{g^{00}} \sqrt{-h}
\!\int\! \frac{d^d\uk}{(2\pi)^d} {\pl\over\pl t}\!
\left[\ln \tpsi(t, \uk)\right] \!\int\! d^d \ux \!\int\! d^d \uy\,
\phi(\ux) e^{i\uk.(\ux-\uy)} \phi(\uy)\right\}
\label{eq:dyn}
\ee
and
\be
N_0(t) = e^{-i\int^t dt E_0(t)}
\label{eq:time1dynamic}
\ee
with
\be
E_0(t) = {1\over2} \sqrt{g_{00}} \sqrt{-h}
\int {d^d\uk}~\widetilde G(\uk; t) \delta^d(0)
\label{eq:energydynamic}
\ee
In this case, however, \(E_0\) does not have a useful interpretation
as an energy.

The analogue of the higher excited states can be found by substituting
for the kernel in the general formulae following eq.(\ref{eq:firstwavecst}), 
where \(f_{\uk}(\ux) = e^{i \uk.\ux} \) 
is an eigenstate of \((\Box_i + m^2 + \xi R)\) with eigenvalue 
\(\omega^2(\uk; t)\).
Again, however, we emphasise that the physical interpretation of
these states is not as straightforward as for a static spacetime
where they are genuine stationary states of the energy
associated with the chosen Hamiltonian evolution.
 
\vspace{0.3cm}
It is important to check the stability of the vacuum state defined by the
time-dependent wave functional (\ref{eq:dyn}). Vacuum stability requires that
\be
\int{\cal D}\phi \Psi_0^*(\phi,t) \Psi_0(\phi,t) = |N_0(t)|^2
\int{\cal D}\phi \psi_0^*(\phi,t) \psi_0(\phi,t)
\label{eq:norm}
\ee
should be time independent. From eqs.(\ref{eq:time1dynamic}) and
(\ref{eq:energydynamic}) we have
\be
  |N_0(t)|^2 =  \exp\{- \half \int  d^d\underline{x} 
  \int \frac{d^d\underline{k}}{(2\pi)^d} 
  \: \ln |\tpsi|^2 \}
\ee
where we have used eq.(\ref{eq:tildeGsoln}) for the kernel. 
Obviously this will in general be time dependent.

The second contribution to eq.(\ref{eq:norm}) is 
\begin{eqnarray}
  \int {\cal D}\phi \: |\psi_0|^2 &\sim&  [\mbox{det}
  ( h \:G_\Re(\ux,\uy;t))]^{-\half} 
  \nonumber \\
  &=&\exp\{ -\half \int  d^d\underline{x} \int \frac{d^d\uk}{(2\pi)^d}
  \ln[ h \:\widetilde{G}_\Re(\uk;t)]\}
\end{eqnarray}
where \(\widetilde{G}_\Re\) is the real part of the (generally complex) kernel.
From eq.(\ref{eq:tildeGsoln}) we see
\be
  \label{eq:realkerneldynamic}
  \widetilde{G}_{\Re}(\underline{k};t) = \frac{i\:W[\tpsi,\tpsistar]}
  {2\sqrt{-g}\:|\tpsi|^2}
 \ee
while for the imaginary part,
\be 
  \widetilde{G}_{\Im}(\underline{k};t) = \frac{-1}
  {2\sqrt{-g}\:|\tpsi|^2} \:\frac{\partial}{\partial t} |\tpsi|^2
\label{eq:imagkerneldynamic}
\ee
where $W[\tpsi,\tpsistar] = \tpsi \, (\frac{\partial}{\partial t} \tpsistar)
- \tpsistar \, (\frac{\partial}{\partial t} \tpsi)$ is the Wronskian of
the solution and its complex conjugate. 
Using eqs.(\ref{eq:ftdynamic}) and (\ref{eq:realkerneldynamic}) 
therefore gives 
\be
  \int {\cal D}\phi \: |\psi_0|^2 \sim 
  \exp\{+ \half \int  d^d\underline{x} \int \frac{d^d\underline{k}}{(2\pi)^d} 
  \: \ln |\tpsi|^2\} 
\ee
up to time-independent terms. These remaining terms are seen to be
time independent using Abel's theorem, viz.~that \(g^{00}\sqrt{-g}W\)
is a constant, where \(W\) is the Wronskian defined above.

The overall time dependence therefore cancels as required,
confirming the stability of the vacuum state in a dynamic spacetime.

\subsection{Conformally static and Robertson-Walker spacetimes}

A straightforward generalisation of the results of sections 3.1 and 3.2
allows us to find the kernel for the class of $(d+1)$-dimensional
spacetimes with metric
\be
  \label{eq:confstaticint1}
  ds^2 = g_{00}(t) dt^2 - a^2(t) d\underline{\s}^2
\ee
where $d\underline{\sigma}^2$ is a $d$-dimensional static
space with metric $\s_{ij}(\ux)$. Let $\s = {\rm det} \s_{ij}$.
 Introducing a rescaled `conformal time' $\eta$, the metric can be written as 
\be
  \label{eq:confstaticint2}
  ds^2  =   C(\eta)\left[d\eta^2 - \s_{ij}(\ux)dx^i dx^j\right]
\ee
The spacetime is therefore conformally static, but with the conformal
scale factor restricted to depend only on time. This class  
includes the important Robertson-Walker spacetimes, which are 
special cases of eq.(\ref{eq:confstaticint1}) where $g_{00}=1$
and $d\underline{\s}^2$ is a 3-dimensional static space of constant
curvature $\kappa = -1, 0, +1$, viz.
\be
  d\underline{\sigma}^2 
  = d\chi^2 + f^2(\chi)[d\theta^2 + \sin^2\theta d\phi^2]
\ee
where 
\be
  f(\chi) = \left\{
    \begin{array}{lcccl}
      \sin \chi&& 0 \leq \chi < 2\pi&&\kappa = +1\\
      \chi&&0 \leq \chi < \infty&&\kappa = 0\\
      \sinh \chi&&0 \leq \chi < \infty&&\kappa = -1
    \end{array}\right.
\ee
For $\kappa = -1, 0, +1$ this metric describes 
hyperbolic (open), flat, and spherical (closed) spaces respectively.

Since the spatial metric is static, general theorems  
guarantee the existence of orthonormal spatial
modes ${\cal Y}_{(\l)}(\ux)$ which play the r\^ole of the Fourier 
functions $\exp i\uk.\ux$ in the previous section.
A full discussion of the solutions of the wave equation for these
spacetimes is given in appendix A.2.

We therefore write the kernel as 
\be
  \label{eq:kernelexpconfstatic}
  G(\underline{x},\underline{y};t) = 
  \int \frac{d\mu(\lambda)}{(2\pi)^d} \:
  \widetilde{G}(\lambda;t)\:
  {\cal Y}_{(\lambda)}(\underline{x})\,
  {\cal Y}^\ast_{(\lambda)}(\underline{y})
\ee
Then, using the solutions ${\cal T}(t,\lambda)$ of
the transformed wave equation (see eq.(\ref{eq:solutionconfstatic})),
we readily find (c.f.~eqs.(\ref{eq:tildeGsoln}) or 
(\ref{eq:kernelsolutiongeneral})) 
\be
  \widetilde{G}(\lambda;t) = -i \frac{\sqrt{\s}}{\sqrt{-g}}\:
  \frac{\partial}{\partial t}\ln {\cal T}(t,\lambda)
\ee
The kernel for a conformally static spacetime with a conformal
factor which is a function of time only is therefore  
\be
  \label{eq:kernelconfstatic}
   G(\underline{x},\underline{y};t) = 
   \frac{-i}{\sqrt{g_{00}}a^{d}}\:
   \int \frac{d\mu(\lambda)}{(2\pi)^d} \,
  \left[\frac{\partial}{\partial t}\ln {\cal T}(t,\lambda)\right]
  {\cal Y}_{(\lambda)}(\underline{x})\,
  {\cal Y}^\ast_{(\lambda)}(\underline{y})
\ee
The interpretation is identical to that for the dynamic metric case.
Again, there is a one-parameter family of vacuum states corresponding to
the freedom in selecting the solution  ${\cal T}(t,\lambda)$ of the
wave equation.

\newpage

\section{Cosmological Model I}
\label{sec-cosm}

To illustrate this formalism, we now consider two simple two-dimensional
`cosmological models' of Robertson-Walker type. We give an exact
description of the vacuum states, emphasising the r\^ole of boundary
conditions, and discuss carefully the phenomenon of particle
creation due to the cosmological expansion. 

The spacetime metric is chosen to be of Robertson-Walker form,
\be
ds^2 = dt^2 - a(t)^2 dx^2
\label{eq:rw}
\ee
Rescaling the time coordinate such that $dt = a(t) d\eta$, the metric may be
rewritten in terms of the `conformal time' $\eta$ as
\be
ds^2 = C(\eta) (d\eta^2 - dx^2)
\ee
The metric is conformal to Minkowski spacetime. 

\begin{figure}[htb]
\centerline{\epsfxsize=3in \epsfbox{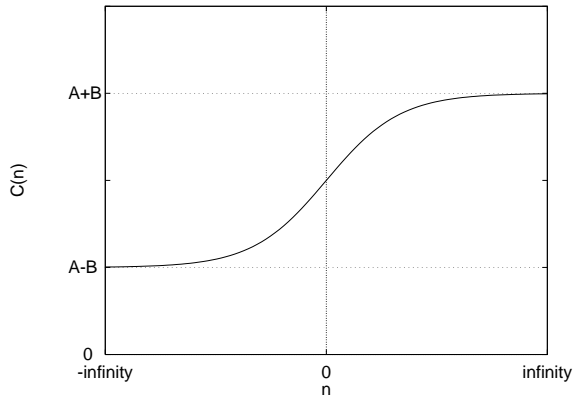}}
\caption{Conformal scale factor for the cosmological model with expansion.}
\label{fig:xfigflat}
\end{figure}

The first model \cite{birrel82} is chosen to have asymptotically 
static regions separated
by a period of expansion (see Fig(\ref{fig:xfigflat})). The conformal scale factor
is chosen to be
\begin{equation}
  \label{eq:conformalfactor}
  C(\eta) = A + B\tanh(\rho \eta)
\end{equation}
where $A,B, \rho$ are constants with $A>B>0$. In the asymptotic IN
($\eta \rightarrow -\infty$) and OUT ($\eta \rightarrow \infty$) regions,
\be
  \label{eq:inoutconffactor}
  C(\eta) \rightarrow A\:\pm \:B \:\:\:\:\:\:\mbox{as}\:\:\eta \rightarrow \pm \infty
\ee
This choice allows explicit analytic solutions for the wave functionals in terms
of hypergeometric functions. 

The metric is in the class we have called `dynamic', so the construction
of the vacuum wave functional is a straightforward application of the
techniques described in section 3.2. The first step is to find the solutions
of the wave equation in the metric specified by eq.(\ref{eq:conformalfactor}).

\subsection{Wave equation}

Taking the Fourier transform,
\be
\psi(\eta,x) = \int_{-\infty}^\infty \frac{dk}{(2\pi)} \:e^{ikx}\:\tpsi(\eta;k)
\ee
the wave equation becomes 
\be
[\partial^2_\eta + (k^2+Am^2) + Bm^2\tanh(\rho \eta)]\tpsi(\eta,k)=0
\ee
Notice that the only occurrence of the conformal scale factor is with the
mass $m$. This is a trivial consequence of the conformal invariance of the
scalar field equation for $m=0$ (we have taken $\xi = 0$ for simplicity).
Obviously any non-Minkowskian behaviour of the vacuum must 
involve the mass.

It is convenient at this point to make the definitions:
\begin{eqnarray}
  \label{eq:frequency}
  \w^2(k;\eta) &=& k^2 + m^2 C(\eta)  \\
  \w^2_{in} &=& k^2 +m^2(A-B)\\
  \w^2_{out} &=& k^2 +m^2(A+B)\\
  \w_\pm &=& \half (\w_{out} \pm \w_{in})
\end{eqnarray}
With the substitution
\be
\tpsi(\eta,k) = e^{-iw_+\eta}\:[2\cosh (\rho \eta)]^{-\frac{iw_-}{\rho}} 
\:\widetilde{\chi}(\eta,k)
\ee
and a change of variable $z=\half[1+\tanh(\rho \eta)]$, 
the wave equation reduces to the hypergeometric equation
\begin{equation}
  \label{eq:hyperde}
  \left\{
    z(1-z)\partial_z^2 + \left[1-\frac{iw_{in}}{\rho} - z(2+2\frac{iw_-}{\rho})\right]\partial_z
    -\frac{iw_-}{\rho} + \frac{w^2_-}{\rho^2}
    \right\}\widetilde{\chi}(z,k) = 0
\end{equation}
The properties of this equation and its solutions are discussed in
\cite{grad80,magnus66}. Here, since $-\infty<\eta<\infty$, we 
are only concerned with solutions in the range $0\leq z \leq 1$.

Because of the singular points at $z=0$ and $z=1$ in the hypergeometric 
equation we have to consider two sets of solutions to cover this range
completely. The first set are valid in the range $0\leq z<1$, 
while the second set are valid  for $0<z\leq1$. The radius of convergence of
the two sets of solutions is strictly less than 1.

Within the radius of convergence of the $z=0$ singular point  there are 
two linearly independent solutions.
We can choose these so that as $z \rightarrow 0$, one, $\tpsi^{in}_+$,
behaves as a positive frequency solution 
while the other, $\tpsi^{in}_-$, becomes a negative frequency solution. 
Explicitly,
\begin{eqnarray}
  \tpsi^{in}_+(\eta,k) &=& e^{-iw_+\eta}[2\cosh (\rho \eta)]^{-\frac{iw_-}{\rho}}
  \times \nonumber \\
  \label{eq:inpossolution}
  &&{_2F_1}\left(\mbox{$\frac{iw_-}{\rho},1+\frac{iw_-}{\rho};1-\frac{iw_{in}}{\rho};z$}\right)\\
  && \nonumber \\
  &\buildrel {\eta \rightarrow -\infty} \over \sim& e^{-iw_{in} \eta}\\
  & & \nonumber \\
  \tpsi^{in}_-(\eta,k) =(\tpsi^{in}_+)^\ast &=&  
e^{-iw_+\eta}[2\cosh (\rho \eta)]^{-\frac{iw_-}{\rho}}
\left[\half(1+\tanh(\rho \eta))\right]^{\frac{iw_{in}}{\rho}} \times \nonumber \\
  \label{eq:innegsolution2}
  & & _2F_1\left(\mbox{$\frac{iw_+}{\rho},1+\frac{iw_+}{\rho};
    1+\frac{iw_{in}}{\rho};z$}\right)\\
  && \nonumber \\
  &\buildrel {\eta \rightarrow -\infty} \over \sim& e^{+iw_{in} \eta}
\end{eqnarray}

The solutions valid around the $z=1$ singular point can similarly be chosen
so that they correspond to positive and negative frequency solutions 
in the limit $z\rightarrow 1$. These are
\begin{eqnarray}
  \label{eq:outpossolution}
  \tpsi^{out}_+(\eta,k) &=& e^{-iw_+\eta}[2\cosh (\rho \eta)]^{-\frac{iw_-}{\rho}} 
  \times \nonumber \\
  & & {_2F_1}\left(\mbox{$\frac{iw_-}{\rho},1+\frac{iw_-}{\rho};
      1+\frac{iw_{out}}{\rho};1-z$}\right)\\
  && \nonumber \\
  &\buildrel {\eta \rightarrow \infty} \over \sim& e^{-iw_{out} \eta}\\
  & & \nonumber \\
  \tpsi^{out}_-(\eta,k) = (\tpsi^{out}_+)^\ast&=& 
  \label{eq:outnegsolution2}
  e^{-iw_+\eta}[2\cosh (\rho \eta)]^{-\frac{iw_-}{\rho}}
  \left[\half(1-\tanh(\rho \eta))\right]^{-\frac{iw_{out}}{\rho}} \times \nonumber \\
  &&{_2F_1}\left(\mbox{$-\frac{iw_+}{\rho},1-\frac{iw_+}{\rho};
    1-\frac{iw_{out}}{\rho};1-z$}\right)\\
  & & \nonumber \\
  &\buildrel {\eta \rightarrow \infty} \over \sim& e^{+iw_{out} \eta}
\end{eqnarray}

\vspace{0.2cm}
In the overlap region $0<z<1$, both sets of solutions are valid and we may 
express one in terms of the other. Specifically, we have  
\be
  \tpsi^{in}_-(\eta,k) = \alpha(k)\tpsi^{out}_- + \beta(k)\tpsi^{out}_+
\label{eq:inout}
\ee
where
\begin{eqnarray}
  \label{eq:alphacosm}
  \alpha(k)&=&
  \frac{\Gamma(1+\frac{iw_{in}}{\rho}) \Gamma(\frac{iw_{out}}{\rho})}
  {\Gamma(\frac{iw_{+}}{\rho})  \Gamma(1+\frac{iw_{+}}{\rho})}\\
  && \nonumber \\
  \label{eq:betacosm}
  \beta(k)&=&
  \frac{\Gamma(1+\frac{iw_{in}}{\rho}) \Gamma(\frac{-iw_{out}}{\rho})}
  {\Gamma(\frac{-iw_{-}}{\rho})  \Gamma(1-\frac{iw_{-}}{\rho})}
\end{eqnarray}
The magnitudes of these two coefficients are
\begin{eqnarray}
  \label{eq:magalphacosm}
  |\alpha(k)|^2 &=& \left(\frac{w_{in}}{w_{out}}\right)
  \:\frac{\sinh^2\left(\frac{\pi w_+}{\rho}\right)}
  {\sinh\left(\frac{\pi w_{out}}{\rho}\right)
    \sinh\left(\frac{\pi w_{in}}{\rho}\right)}\\
  && \nonumber \\
  \label{eq:magbetacosm}
  |\beta(k)|^2 &=& \left(\frac{w_{in}}{w_{out}}\right)
  \:\frac{\sinh^2
    \left(\frac{\pi w_-}{\rho}\right)}
  {\sinh\left(\frac{\pi w_{out}}{\rho}\right)
    \sinh\left(\frac{\pi w_{in}}{\rho}\right)}
\end{eqnarray}
and satisfy
\begin{equation}
  \label{eq:aaminusbbcosm}
  |\alpha(k)|^2 - |\beta(k)|^2 = \left(\frac{w_{in}}{w_{out}}\right)
\end{equation}

Notice that the solution which is positive frequency in the $z\rightarrow 0$ 
IN region is a mixture of the solutions of both positive and negative
frequencies in the $z\rightarrow 1$ OUT region. As we shall see, this mixing 
is at the heart of the `particle creation' interpretation. The $\a$ and $\b$
coefficients above are just the Bogoliubov coefficients in conventional
treatments \cite{bernard77,birrel82}.

\subsection{Vacuum wave functional}

The vacuum wave functional is given by the standard formula
(\ref{eq:kerndynamic})  for a dynamic spacetime. That is,
\be
  G(x,y;\eta) = \int_{-\infty}^{\infty} \frac{dk}{(2\pi)} 
  e^{ik(x-y)}\widetilde{G}(k;\eta)
\ee
where
\be
  \widetilde{G}(k;\eta) = -\frac{i}{C(\eta)}\frac{\partial}{\partial \eta}
   \ln [ a \tpsi_+^{in}(\eta,k) + b \tpsi_-^{in}(\eta,k)]
\ee

To specify the vacuum beyond this one-parameter ambiguity, we need to
impose a boundary condition. We choose this so that in the IN region, the
wave functional becomes the usual positive frequency Minkowski vacuum
\be
\psi_0^{IN}[\tphi] = \exp \biggl\{ -{1\over2} \int_{-\infty}^{\infty}
{dk\over{2\p}} \w_{in} |\tphi(k)|^2 \biggr\}
\label{eq:invac}
\ee
This implies we choose $a=0, b=1$ above, i.e.~use the 
negative frequency solution of the wave equation in the dynamic kernel.

Explicitly, 
\begin{eqnarray}
  \label{eq:inkernel}
  \widetilde{G}(k;\eta) &=& -\frac{i}{C(\eta)}\frac{\partial}{\partial \eta}
  [\ln \tpsi^{in}_-(\eta,k)] \\
   &=&{-1\over C(\eta)} \biggl\{\w_+ + \w_-\tanh(\r\eta)
    - \w_{in}[1-\tanh(\r\eta)] \times  \nonumber \\
      &&~~~\left[\frac
        {_2F_1\left(\mbox{$\frac{i}{\rho}\w_+,1+\frac{i}{\rho}\w_+;
    \frac{i\w_{in}}{\rho};z$}\right)}{
  _2F_1\left(\mbox{$\frac{i}{\rho}\w_+,1+\frac{i}{\rho}\w_+;
    1+\frac{i\w_{in}}{\rho};z$}\right)}\right]\biggr\}
\end{eqnarray}
which in the asymptotic limit is simply
\be
  \widetilde{G}(k;\eta \rightarrow -\infty) = \frac{\w_{in}}{(A-B)}
  \label{eq:inftkernelcosm}
\ee
as required to reproduce eq.(\ref{eq:invac}).
The complete vacuum wave functional is therefore
\be
\Psi_0[\tphi,\eta] = N_0(\eta) \psi_0[\tphi, \eta]
\ee
where
\be
\psi_0[\tphi,\eta] = \exp \biggl\{-{1\over2} C(\eta) \int_{-\infty}^\infty
{dk\over2\p}
\widetilde{G}(k;\eta) |\tphi(k)|^2
\ee
and
\be
N_0(\eta) = \exp \biggl\{ -{i\over2} \int^{\eta} d\eta  C(\eta)
\int^\infty_{-\infty} dk  \widetilde{G}(k;\eta) \delta(0)  \biggr\}
\ee

This is the full analytic expression, valid for all values of the time $\eta$, 
for the vacuum state chosen to 
satisfy a positive frequency boundary condition in the IN region.
It is stable for all times, the proof being precisely that given in section 3
for a general dynamic metric. 
 
The next step is to examine the behaviour of this state in the asymptotic
future OUT region. For this, we use eq.(\ref{eq:inout}) to reexpress the
kernel in terms of the wave equation solutions $\tpsi_+^{out}$ and
$\tpsi_-^{out}$ which have simple asymptotic behaviour as $z\rightarrow 1$.
This gives
\be
  \widetilde{G}(k;\eta) = \frac{-i}{C(\eta)}\left[\frac
    {{\alpha}\frac{\partial}{\partial \eta}(\tpsi^{out}_-)
      +{\beta}\frac{\partial}{\partial \eta}(\tpsi^{out}_+)}
    {{\alpha}(\tpsi^{out}_-) + {\beta}(\tpsi^{out}_+)}\right]
\ee
As $\eta \rightarrow \infty$ this function does not tend to a fixed value 
but rather to a limit circle in the complex plane. We find
\be
  \label{eq:inoutkernelcosm}
  \widetilde{G}(k;\eta\rightarrow \infty)  =
  \frac{\w_{out}}{(A+B)} \left[
    \frac{1 -{\gamma} e^{-2i\w_{out}\eta}}
    {1 +{\gamma} e^{-2i\w_{out}\eta}}\right]
\ee
where ${\gamma}=\frac{{\beta}}{{\alpha}}$. 
The field-dependent part of the wave functional is therefore
\be
  \label{eq:inoutwavefnalcosm}
    \psi_0[\tphi,\eta\rightarrow\infty] =
  \exp \left\{-\half  \int_{-\infty}^{\infty} \frac{dk}{2\pi} 
    \:\w_{out} \left[
    \frac{1 - {\gamma} e^{-2i\w_{out}\eta}}
    {1 + {\gamma} e^{-2i\w_{out}\eta}}\right]
    \: |\tphi(k)|^2\right\}
\label{eq:asympvac}
\ee

\subsection{Vacuum states and particle creation}

We now discuss the interpretation of this vacuum state and in particular  
the phenomenon of particle creation.
\vspace{0.3cm}
First, notice that eq.(\ref{eq:asympvac}) is {\it not} the standard
Minkowski vacuum corresponding to the asymptotic limit of the metric, 
viz.
\be
\label{eq:outminkvac}
     \psi_0^{OUT}[\tphi] =
  \exp \left\{-\half  \int_{-\infty}^{\infty} \frac{dk}{2\pi} 
    \:\w_{out}     \: |\tphi(k)|^2\right\}
\ee 
In fact, this is the asymptotic limit of a different state 
$\Psi_0^{\prime}[\tphi,\eta]$ chosen from the one-parameter family
of vacuum solutions to the Schr\"odinger equation, specified by the kernel 
(c.f.~eq.(\ref{eq:inkernel}))
\be
\widetilde{G}^{\prime}(k;\eta) = -{i\over C(\eta)} {\partial\over\partial\eta}
[\ln \tpsi_-^{out}(\eta,k)]
\ee

Completeness of the solutions to the Schr\"odinger equation means
that any state can be expressed as a superposition of one of the
one-parameter family of vacuum solutions and the excited states
built from it. In particular, we can express the chosen
vacuum state $\Psi_0[\tphi,\eta]$ as a linear combination of the state 
$\Psi_0^{\prime}[\tphi,\eta]$ and its excited states. In the OUT region, this 
amounts to expressing the Gaussian state $\psi_0[\tphi,\eta\rightarrow\infty]$
of eq.(\ref{eq:inoutwavefnalcosm}) in terms of the excited states of the 
Minkowski state $\psi_0^{OUT}[\tphi]$. 

In the limit $\eta\rightarrow\infty$, these excitations are standard Minkowski
particle states and we can characterise them by their particle number and
momenta. The problem therefore reduces to the one solved in section 2.1,
i.e.~expressing a general, time-dependent Gaussian state in Minkowski 
spacetime in terms of a superposition of many-particle states. 
The time-dependent kernel $\Omega$ is read off from 
eq.(\ref{eq:inoutwavefnalcosm}), viz.
\be
\Omega(k;\eta) = \w_{out}  \left[
    \frac{1 -{\gamma} e^{-2i\w_{out}\eta}}
    {1 +{\gamma} e^{-2i\w_{out}\eta}}\right]
\label{eq:omega}
\ee
Transcribing results directly from section 2.1, we find that the expectation
value of the number density of particles with momentum $k$ in 
the state $\Psi_0[\tphi,\eta\rightarrow\infty]$ is 
\be 
\langle \Psi_0[\tphi,\eta\rightarrow\infty] | {\cal N}(k) | 
\Psi_0[\tphi,\eta\rightarrow\infty] \rangle 
  = \frac{(\w_{out}-\Omega)(\w_{out}-{\Omega}^*)}
{2 \w_{out} (\Omega +{\Omega}^*)}
\ee
Remarkably, despite the residual time-dependence in $\Omega$,
this expectation value is independent of $\eta$.
Explicitly,
\begin{eqnarray}
\label{eq:evnuoperatorcosm}
\langle {\cal N}(k) \rangle &=&  \frac{|\gamma|^2}{1-|\gamma|^2}
 ~=~ \frac{\w_{out}}{\w_{in}} \: |\beta|^2 \\
   &=&      \:\frac{\sinh^2
    \left(\frac{\pi \w_-}{\rho}\right)}
  {\sinh\left(\frac{\pi \w_{out}}{\rho}\right)
    \sinh\left(\frac{\pi \w_{in}}{\rho}\right)}
\end{eqnarray}
This is the Schr\"odinger picture derivation of the traditional result 
on particle creation in this expanding universe model. Notice that
the particle number density is proportional to $|\b|^2$, the 
Bogoliubov coefficient in the conventional treatment, which is only
non-zero if there is a mixing between the positive and negative frequency 
solutions of the wave equation in the respective asymptotic regions.
As expected, $\langle {\cal N}(k) \rangle$ also vanishes in the 
conformal limit $m\rightarrow 0$.

\vspace{0.3cm}
It is also illuminating to evaluate the expectation value of the
energy momentum tensor in these states. 
In particular, $\langle \Psi_0 | T_{00} | \Psi_0 \rangle$ evaluated in 
the asymptotic IN and OUT regions measures the time-independent energy
eigenvalue of the state $\Psi_0$. In the intermediate region, of course,
$\Psi_0$, like all the other `vacuum' solutions, is not an energy eigenstate. 
With the definition of $T_{00}$ implicit in eq.(\ref{eq:swfecst}) 
(recall that $H = \int d^d\ux ~\sqrt{-g} g^{00} T_{00}$), we find
\be 
\langle \Psi_0[\tphi,\eta] |T_{00}|\Psi_0[\tphi,\eta]\rangle
= {1\over2}C(\eta) \int_{-\infty}^\infty {dk\over 2\p} \widetilde{G}(k;\eta) 
\biggl[1 - {1\over 2 \widetilde{G} \widetilde{G}_{\Re}} \Bigl(
\widetilde{G}^2 - {\w^2\over C(\eta)^2}\Bigr) \biggr]
\label{eq:energyvev}
\ee
where we have normalised $\Psi_0$ such that 
$\langle \Psi_0 | \Psi_0\rangle = 1$.

In the IN region, we simply have
\be
\langle \Psi_0 |T_{00}|\Psi_0\rangle_{IN} = {1\over2} \int_{-\infty}^\infty
{dk\over2\p}~ \w_{in}
\ee
This is just the usual Minkowski zero-point energy density. On the other hand,
substituting (\ref{eq:inoutkernelcosm}) into eq.(\ref{eq:energyvev}) 
we find, after some calculation,
\be
\langle \Psi_0 |T_{00}|\Psi_0\rangle_{OUT} = 
\int_{-\infty}^\infty {dk\over2\p}~ \w_{out}~\Bigl( \langle {\cal N}(k)\rangle + 
{1\over2} \Bigr)
\label{eq:energy}
\ee
Notice that the $\eta$ dependence from the kernel 
$\widetilde{G}(k;\eta\rt\infty)$ has again cancelled out.
$\Psi_0[\tphi,\eta\rt\infty]$ is therefore not the minimum energy state
in the Minkowski OUT region.
Rather, eq.(\ref{eq:energy}) describes the energy of a many-particle
state with $\langle {\cal N}(k)\rangle$ particles of energy 
$\w_{out}(k)$ as described above.

\vspace{0.3cm}
The Schr\"odinger picture therefore brings a particular clarity to the
issue of particle creation in this model. It is evident that there is no state
which satisfies all the properties of a standard Minkowski vacuum state. 
As we have repeatedly emphasised, there is a one-parameter family of 
Gaussian `vacuum' solutions to the wave functional equation for this
class of dynamic spacetimes. If we choose to consider the state
$\Psi_0[\tphi,\eta]$ which in the asymptotic past is the Minkowski vacuum
state $\psi_0^{IN}[\tphi]$ and follow its evolution, we find that in the
asymptotic future OUT region it is described by the time-dependent
kernel (\ref{eq:inoutkernelcosm}). At this point, there is really no more
to be said -- that is a complete description of the state. It is stable and
any physical quantity may be evaluated by taking appropriate
operator matrix elements. The remaining question is simply to 
describe what the state $\Psi_0[\tphi,\eta]$ looks like.
As we have shown above, in the OUT region where the spacetime 
is again asymptotically Minkowskian, it is precisely a superposition of 
many-particle states built from the minimum energy Minkowski vacuum state
$\psi_0^{OUT}[\tphi]$. In the asymptotic OUT region, therefore, it 
is true to say that particles have been created as a result of the 
expansion. For any intermediate value of the time, $\Psi_0[\tphi,\eta]$
continues to provide a complete description of the state, although
in this case there is no simple interpretation in terms of Minkowski particles.

\newpage

\section{Cosmological Model II}

In this section, we consider a second cosmological model of the same 
type \cite{grib94}, this time with a conformal scale factor
\be
C(\eta) = {A^2\over \cosh^2(\r\eta)}
\ee
\begin{figure}[htb]
\centerline{\epsfxsize=3in \epsfbox{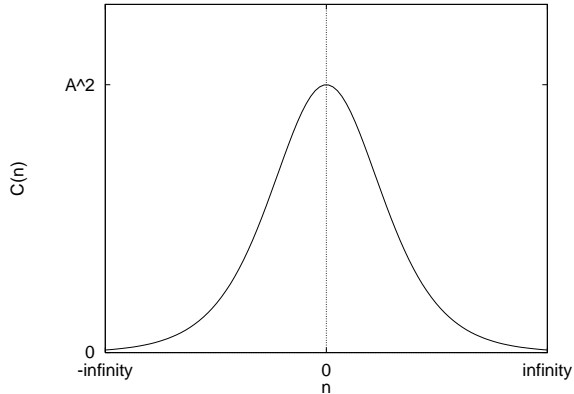}}
\caption{Conformal scale factor for the recollapse model.}
\label{fig:xfigpotential}
\end{figure}
This is shown in Fig(\ref{fig:xfigpotential}).
This describes a universe which expands exponentially from an initial
singularity before eventually recollapsing. (Since $C(\eta)\rt 0$
as $\eta\rt \pm\infty$, the size of the universe tends to zero in the
past and future asymptotic limits.) Notice that although 
the universe is infinitely long-lived in terms of the conformal time $\eta$,
the range of the Robertson-Walker time coordinate $t$ in 
eq.(\ref{eq:rw}) is only from $t=0$ to $t= T = A\p/\r$.
In fact, solving the relation $dt^2 = C(\eta) d\eta^2$, we find
$a(t) = A \sin{\r t /A}$, representing an oscillating universe
of which we are considering only the first cycle.

The analysis of this model follows closely that of section 4.
Again we have two asymptotically flat regions in which we can
compare the vacuum solutions of the dynamic spacetime Schr\"odinger
equation with the equivalent Minkowski vacua. With some additional
subtleties, the results are similar and we again find particle creation
occurring as a result of the gravitational expansion and recontraction.

\subsection{Wave equation}

In this case, the Fourier transformed wave equation is 
\be
[ \partial^2_\eta + k^2 +A^2 m^2 \cosh^{-2}(\r\eta)]\tpsi(\eta,k)=0
\ee
Changing variables from $\eta$ to $z$ as before and making the substitution
\be
\tpsi(\eta,k) = [2\cosh (\r \eta)]^{-\frac{ik}{\rho}} 
\:\widetilde{\chi}(\eta,k)
\ee
the wave equation reduces to the hypergeometric equation 
\cite{grad80, magnus66}
\be
  \label{eq:hyperde2}
  \left\{ z(1-z)\partial_z^2 + \left[1+\frac{ik}{\rho} - 
      z(2+2\frac{ik}{\rho})\right]\partial_z
    +\left[\frac{k^2}{\rho^2} - \frac{ik}{\rho} + \frac{A^2 m^2}{\rho^2}
      \right]
    \right\}\widetilde{\chi}(z,k) = 0
\ee
Again, there are two sets of solutions of interest, valid in the neighbourhood
of the singular points $z=0$ and $z=1$ with a radius of convergence 
strictly less than 1. Choosing a basis of asymptotically
positive or negative frequency solutions, we have 
\begin{eqnarray}
  \label{eq:innegsolutionII}
  \tpsi^{in}_-(\eta,k) &=& [2\cosh (\rho \eta)]^{-\frac{ik}{\rho}}
  {_2F_1}\left(\mbox{$\frac{ik}{\rho}-\sigma,1+\frac{ik}{\rho}+\sigma
      ;1+\frac{ik}{\rho};z$}\right) \nonumber \\
  &&  \\
  &\buildrel {\eta \rightarrow -\infty} \over \sim& e^{+ i k \eta} \nonumber \\
  & & \nonumber \\
  \label{eq:inpossolutionII}
  \tpsi^{in}_+(\eta,k) = (\tpsi^{in}_-)^* 
&=& [2\cosh (\rho \eta)]^{-\frac{ik}{\rho}}
  [z]^{-\frac{ik}{\rho}}
  {_2F_1}\left(\mbox{$-\sigma,1+\sigma
      ;1-\frac{ik}{\rho};z$}\right)  \nonumber \\
  && \\
  &\buildrel {\eta \rightarrow -\infty} \over \sim& e^{- i k \eta} \nonumber 
\end{eqnarray}
and
\begin{eqnarray}
 \label{eq:outnegsolutionII}
  \tpsi^{out}_-(\eta,k) &=& \! \! [2\cosh (\rho \eta)]^{-\frac{ik}{\rho}}
  [1-z]^{-\frac{ik}{\rho}}
  {_2F_1}\left(\mbox{$-\sigma,1+\sigma
      ;1-\frac{ik}{\rho};1-z$}\right)  \nonumber \\
  &&  \\
  &\buildrel {\eta \rightarrow \infty} \over \sim& e^{+ i k \eta} \nonumber \\
  &&\nonumber \\
  \label{eq:outpossolutionII}
  \tpsi^{out}_+(\eta,k)  = (\tpsi^{out}_-)^*
&=& \! \! [2\cosh (\rho \eta)]^{-\frac{ik}{\rho}}
  {_2F_1}\left(\mbox{$\frac{ik}{\rho}-\sigma,1+\frac{ik}{\rho}+\sigma
      ;1+\frac{ik}{\rho};1-z$}\right)  \nonumber \\
  &&  \\
  &\buildrel {\eta \rightarrow \infty} \over \sim& e^{- i k \eta} \nonumber 
\end{eqnarray}
where
\be
  \label{eq:sigma}
  \sigma = -\half \ \pm \ \half \sqrt{1+\mbox{$\frac{4 A^2 m^2}{\rho^2}$}}
\ee

Notice that these solutions are independent of the mass $m$ in the
asymptotic limit. The mass decouples when the size of the universe
shrinks beyond the scale of $m^{-1}$.

In the intermediate region $0<z<1$, we can express these solutions
in terms of each other. In particular, we will need the relation
\be
  \tpsi^{in}_-(\eta,k) = \alpha(k) \tpsi^{out}_- + \beta(k) \tpsi^{out}_+
\ee
where
\begin{eqnarray}
  \alpha(k) &=& \frac{\Gamma(1+\frac{ik}{\rho}) \Gamma(\frac{ik}{\rho})}
  {\Gamma(\frac{ik}{\rho}-\sigma)\Gamma(1+\frac{ik}{\rho}+\sigma)} \\
  && \nonumber \\
  \beta(k) &=& \frac{\Gamma(1+\frac{ik}{\rho}) \Gamma(-\frac{ik}{\rho})}
  {\Gamma(1+\sigma)\Gamma(-\sigma)}
\end{eqnarray}
The magnitudes of these coefficients are
\begin{eqnarray}
  \label{eq:magalphacosmII}
  |\alpha(k)|^2 &=& 
  \frac{\sin^2(\pi \sigma) \cosh^2(\pi k) +\cos^2(\pi \sigma) \sinh^2(\pi k)}
  {\sinh^2(\pi k)}\\
  && \nonumber \\
  \label{eq:magbetacosmII}
  |\beta(k)|^2 &=& \frac{\sin^2(\pi \sigma)}{\sinh^2(\pi k)}
\end{eqnarray}
and satisfy
\be
  \label{eq:aaminusbbcosmII}
  |\alpha(k)|^2 - |\beta(k)|^2 = 1
\ee

\subsection{Vacuum Wave Functional and Particle Creation}

The construction of the vacuum wave functional is identical to 
section 4. Choosing boundary conditions such that the vacuum wave
functional $\Psi_0[\tphi,\eta]$ becomes the standard Minkowski vacuum
in the asymptotic IN region, viz.
\be
\label{eq:inoutwavefnalpot}
  \psi^{IN}_0[\tphi] = 
   \exp \left\{-\half  \int_{-\infty}^{\infty} \frac{dk}{(2\pi)} \: k \: 
    \: |\tphi(k)|^2\right\}
\ee
fixes the one-parameter freedom of
vacuum states. This choice determines the kernel,
\be
  \widetilde{G}(k;\eta) = -\frac{i}{C(\eta)}
  \frac{\partial}{\partial \eta}[ \ln \tpsi^{in}_-(\eta,k)]
\label{eq:model2ker}
\ee
Explicitly, 
\begin{eqnarray}
  \widetilde{G}(k;\eta)&=& - \frac{i}{C(\eta)}\biggl\{
    -i k \tanh(\rho \eta) + \half[1-\tanh^2(\rho \eta)] \times
    \nonumber \\      &&
     \frac{(\frac{ik}{\rho}-\sigma)(1+\frac{ik}{\rho}+\sigma)}
     {(1+\frac{ik}{\rho})} ~
  \frac{{_2F_1}\left(1+\frac{ik}{\rho}-\sigma,2+\frac{ik}{\rho}
  +\sigma  ;2+\frac{ik}{\rho};z\right)}
   {{_2F_1}\left(\frac{ik}{\rho}-\sigma,1+\frac{ik}{\rho}+\sigma
      ;1+\frac{ik}{\rho};z\right)}    \biggr\}
\end{eqnarray}

Rewriting eq.(\ref{eq:model2ker}) in terms of the wave equation solutions
$\tpsi^{out}_{\pm}$ valid in the OUT region, we have 
\be
  \widetilde{G}(k;\eta) = \frac{-i}{C(\eta)}\left[\frac
    {{\alpha}\frac{\partial}{\partial \eta}(\tpsi^{out}_-)
      + {\beta}\frac{\partial}{\partial \eta}(\tpsi^{out}_+)}
    {{\alpha}(\tpsi^{out}_-) + {\beta}(\tpsi^{out}_+)}\right]
\ee
Then, taking the asymptotic future limit $\eta\rt\infty$, we find
the following expression for the $\tphi$ dependent part of the
vacuum wave functional:
\be
  \label{eq:inoutwavefnalcosmII}
    \psi_0[\tphi,\eta\rt\infty]  =
  \exp \left\{-\half  \int_{-\infty}^{\infty} \frac{dk}{(2\pi)} 
    \:k \left[\frac{1 - \gamma (1-z)^{\frac{ik}{\rho}} }
    {1 + \gamma (1-z)^{\frac{ik}{\rho}} } \right]
    \: |\tphi(k)|^2\right\}
\ee
where $\gamma = {\beta\over\alpha}$.
This is to be compared with the equivalent Minkowski vacuum in the
OUT region,
\be
  \psi^{OUT}_0[\tphi] = 
   \exp \left\{-\half  \int_{-\infty}^{\infty} \frac{dk}{(2\pi)} \: k \: 
    \: |\tphi(k)|^2\right\}
\ee
which is of course identical to $\psi^{IN}_0$ in this case.
The same argument as in section 4.3 now shows that the asymptotic vacuum 
state (\ref{eq:inoutwavefnalcosmII}) is a linear superposition of
many-particle states built from the Minkowski vacuum $\psi_0^{OUT}$.
The expectation value of the particle number 
density is
\begin{eqnarray}
\langle \Psi_0[\tphi,\eta\rightarrow\infty] | {\cal N}(k) | 
\Psi_0[\tphi,\eta\rightarrow\infty] \rangle 
&=& {|\c|^2\over 1 - |\c|^2} \nonumber \\
&=&{\sin^2(\p\s) \over \sinh^2(\p k)} 
\label{eq:partno}
\end{eqnarray}
in agreement with the canonical approach \cite{grib94}.
Once again, therefore, we see that the initial Minkowski vacuum has
evolved in the asymptotic future to a state described by a 
Gaussian wave functional with a time-dependent kernel, 
which may be interpreted as a many-particle state. 
We conclude that despite the recontraction following the
initial exponential expansion, particles have been created in this
cosmological model as a result of the time-dependence of the metric. 

A remarkable feature of this model, however, is that for certain values
of the parameters, there is {\it no} particle creation. From eq.(\ref{eq:partno})
we see that $\langle {\cal N}(k)\rangle$ vanishes when
$\s$ is an integer $n$, i.e.
\be
{Am\over\r} = \sqrt{n(n+1)}
\ee
Equivalently, there is no particle creation when the cycle period
of the Robertson-Walker metric and the mass are related by
\be
T = {1\over m} \sqrt{n(n+1)}~ \p
\ee 

\newpage 

\section{Outlook}

These simple models already illustrate some of the power of the 
Schr\"odinger picture techniques to describe vacuum states in 
spacetimes with time-dependent metrics. As we have seen in the 
discussion of cosmological particle creation, there is a clear advantage 
in such spacetimes of representing the vacuum state by a Gaussian
wave functional, entirely characterised by a simple kernel function,
rather than the usual Fock space description. This approach 
also makes clear that it is only in the very special case of certain
static spacetimes that `vacuum' states exist which share all the 
properties associated with the usual Minkowski vacuum.

In the companion paper, we extend this investigation of the 
Schr\"odinger picture to simple model spacetimes with boundaries.
In this case, even when the spacetime is static, the Unruh effect 
sharpens the question of defining a vacuum state by requiring us to 
identify which, if any, class of observers (modelled in the Schr\"odinger
picture by the choice of foliation\footnote{This statement requires
careful qualification, given the fact that for specified boundary conditions,
physical quantities are independent of the choice of foliation.
See ref.(\cite{me96}) for an analysis of the subtleties associated with
observer and foliation dependence.}) perceives the chosen state to
resemble a Minkowski vacuum.

A complete understanding of all these issues is a necessary precursor to
describing vacuum states and particle creation in Schwarzschild-Kruskal
black hole spacetimes, which comprise both static and dynamic regions 
separated by boundaries which are event horizons.

\section*{Acknowledgements}

One of us (GMS) would like to thank Prof.~J--M.~Leinaas for extensive
discussions and hospitality at the University of Oslo and the Norwegian
Academy of Sciences. We are both grateful to Dr.~W.~Perkins 
for many helpful discussions. DVL acknowledges the financial support
of a PPARC research studentship. 

\newpage
\appendix

\section{Wave Equation}
\label{sec-wave}

The wave equation for a massive scalar field is
\be
  (\Box + m^2 + \xi R)\psi(x)=0
\ee
where the Laplacian operator is
\be
  \Box = \frac{1}{\sqrt{-g}}\pl_\m ( g^{\m \n} \sqrt{-g}
  \pl _{\n})
\ee
and \(g = {\rm det }(g_{\m \n})\). The wave equation can be rewritten
as
\be
  (\Box_0 + \Box_i + m^2 + \xi R)\psi(x)=0
\ee
where
\be
  \Box_i = \frac{1}{\sqrt{-g}}\pl_i (g^{ij} \sqrt{-g} \pl_j)
\ee
\be
  \Box_0 = \frac{1}{\sqrt{-g}}\pl_0 (g^{00} \sqrt{-g} \pl_{0})
\ee
in coordinates where the metric components \(g_{0i}\) vanish.

\subsection{Static and dynamic spacetimes}

If the metric is independent of a coordinate, then there exists a
Killing vector associated with the corresponding translation symmetry.
We use this to introduce convenient Fourier transforms for the
special cases of static and dynamic spacetimes.
 
\vspace{0.5cm}
In the static case, because of the translational invariance of the
wave equation with respect to \(t\), the solution may be written as
\be
  \psi(x) = \int_{-\infty}^\infty 
\frac{dw}{(2\pi)} ~ e^{-i\w t}\tpsi(\w,\ux)
\label{eq:fieldstatic}
\ee
where
\be
  (\Box_i - g^{00}\w^2 + m^2 + \xi R) ~\tpsi(\w,\ux)=0
\ee
This equation has degenerate solutions labelled by a discrete or
continuous set of quantum numbers \((\l)\), which generalise
the momentum \(\uk\) in flat space.
 
An extensive discussion of the general properties of such solutions
in static spacetimes has been given by Fulling \cite{fulling89,fulling73}.
They satisfy the orthonormality and completeness relations
\be
  \int \frac{d^d\ux}{(2\pi)^d} \sqrt{-h_x}  \sqrt{g_x^{00}}~
  \tpsistar_{(\l)}(\w,\ux) ~\tpsi_{(\r)}(\w,\ux) ~
  = ~\delta(\l,\r)
\label{eq:orthstatic}
\ee
and
\be
  \int \frac{d\m(\l)}{(2\pi)^d} ~  \tpsistar_{(\l)}(\w,\ux) 
  ~\tpsi_{(\l)}(\w,\uy) ~
  =~ \sqrt{g_{00}^x} \delta^d(\ux,\uy)
\label{eq:compstatic}
\ee
where the measure \(\m(\l)\) and delta function \(\d(\l,\r)\)
are such that
\[
\int d\m(\l)
 ~\delta(\l,\r) ~\widetilde{f}(\l)~ =~ \widetilde{f}(\r)
\]
 
\vspace{0.5cm}
In the dynamic case, because of the translational invariance of the
wave equation with respect to \(x^i\), the solution may be written as
\be
\psi(x) = \int \frac{d^d\uk}{(2\pi)^d} ~ e^{i\uk . \ux} ~
\tpsi (t, \uk)
\label{eq:fielddynamic}
\ee
where \(\uk . \ux \equiv g^i{}_j   k_i x^j \),
the measure is defined in terms of the covariant vector components
as \(d^d\uk = dk_1 \ldots dk_d\)
and \(\tpsi(t,\uk)\) satisfies
\be
  (\Box_0 - g^{ij}k_i k_j + m^2 + \xi R)\tpsi(t, \uk) =0
\label{eq:wavedynamic}
\ee

\subsection{Conformally static and Robertson-Walker spacetimes}

Now consider spacetimes with a conformally static metric
of the form (\ref{eq:confstaticint1}). 
The Laplacian $\Box_0$ becomes
\be
  \Box_0 = \sqrt{g^{00}} a^{-d} \partial_0(\sqrt{g^{00}}a^{d}\partial_0)
\ee
and all dependence on the spatial coordinates drops out.
The Laplacian $\Box_i$ is 
\begin{eqnarray}
  \Box_i &=& \frac{-1}{a^2 \sqrt{\s}} \partial_i (
  \s^{ij} \sqrt{\s}  \partial_j)  \nonumber \\
  &=& -a^{-2} \Box_d
\end{eqnarray}
where $\Box_d$ is the Laplacian on the $d$-dimensional static 
space with metric $\s_{ij}$. The wave equation takes the form
\be
  [a^2 \Box_0 - \Box_d + a^2(m^2 + \xi R)]\psi(x)=0
\ee

The solutions factorise and we find
\be
  \psi(t,\ux) = {\cal Y}_{(\l)}(\ux) {\cal T}(t,\l)
  \label{eq:solutionconfstatic}
\ee
where ${\cal Y}_{(\l)}(\ux)$ are eigenfunctions of 
$\Box_d$ with eigenvalues $K^2(\l)$, i.e. 
\be
  \label{eq:eigenconfstatic}
  \Box_d{\cal Y}_{(\l)}(\ux) = - K^2(\l) {\cal Y}_{(\l)}(\ux)
\ee
and ${\cal T}(t,\l)$ is a solution of the transformed wave equation
\be
  [a^2 \Box_0 + K^2(\l) + a^2(m^2 + \xi R)]{\cal T}(t,\l)=0
\ee
All the analysis above for static spacetimes applies to
eq.(\ref{eq:eigenconfstatic}). General theorems assure the existence 
of a complete, orthonormal set of solutions ${\cal Y}_{(\l)}(\ux)$
satisfying 
\begin{eqnarray}
  \label{eq:orthconfstatic}
  \int \frac{d^d\underline{x}}{(2\pi)^d} \,\sqrt{\s}\:
  {\cal Y}_{(\lambda)}(\underline{x})\,
  {\cal Y}^\ast_{(\rho)}(\underline{x})\,
  &=&\delta(\lambda,\rho) \\
  \int \frac{d\mu(\lambda)}{(2\pi)^d} \,
  {\cal Y}_{(\lambda)}(\underline{x})\,
  {\cal Y}^\ast_{(\lambda)}(\underline{z})\,
  &=&\frac{\delta^d(\underline{x}-\underline{z})}{\sqrt{\s}}
  \label{eq:compconfstatic}
\end{eqnarray}

Robertson-Walker spacetimes are a special case of 
this class of conformally static metrics.
The solutions of the wave equation for $\kappa = -1,0,+1$
are therefore
\be
  \psi(t,\ux) = {\cal T}(t,\l)\:{\cal Y}_{\l}(\chi,\theta,\phi)
\ee
where ${\cal Y}_{\l}$ are eigenfunctions of the 3-dimensional
Laplacian $\Box_3$,  
\be
  \Box_3 {\cal Y}_{\l}(\chi,\theta,\phi) = -K^2(\l)
  {\cal Y}_{\l}(\chi,\theta,\phi)
\ee
and ${\cal T}(t,\l)$ satisfies
\be
  [a^{-1}\partial_t(a^{3}\partial_t) +
    K^2(\l) + a^2(m^2 +\xi R)]{\cal T}(t,\l) = 0
    \label{eq:waverw}
\ee
Here, $K^2(\l) = k^2 -\kappa$, where $(\l) = (k,j,m)$ and the 
eigenfunctions ${\cal Y}_{(\l)}(\ux)$,\cite{parker74,lifshitz63} 
are explicitly 
\be
  \label{eq:rwwavesolutions}
  {\cal Y}_{\bl}(\ux) = 
  \left\{\begin{array}{lccr}
      (2\pi)^{\sfrac{3}{2}} \Pi_{kj}^{(-)}(\chi) \:Y^m_j(\theta,\phi)
      &&&\kappa= -  1\\
      & & \\
      (2\pi)^{\sfrac{3}{2}}\sqrt{\frac{k}{\chi}}
      \:J_{j+\half}(k\chi)\:Y^m_j(\theta,\phi)&&&\kappa=0\\
      & & \\
      (2\pi)^{\sfrac{3}{2}} \Pi_{kj}^{(+)}(\chi) \:Y^m_j(\theta,\phi)
      &&&\kappa= + 1
    \end{array}\right\}
\ee
The $Y^m_j$ are spherical harmonics 
and the function $\Pi^{(-)}$ is defined by
\be
  \Pi_{kj}^{(-)}(\chi) = [\half \pi k^2(k^2+1)\dots(k^2+j^2)]^{-\half}
  \sinh^j \chi
  \left(
    \frac{d^{j+1}\cos (k\chi)}{d(\cosh \chi)^{j+1}}
  \right)
\ee
The function $\Pi_{kj}^{(+)}$ is obtained from $\Pi_{kj}^{(-)}$ by replacing $k$
with $-ik$ and $\chi$ with $-i\chi$. Alternative forms of this 
decomposition are given in \cite{grib94,ford76}.

The orthonormality and completeness relations 
are given by eqs.(\ref{eq:orthconfstatic}) and 
(\ref{eq:compconfstatic}), viz.
\begin{eqnarray}
  \int \frac{d^3\underline{x}}{(2\pi)^3} \,\sqrt{\s}\:
  {\cal Y}_{(\lambda)}(\underline{x})\,
  {\cal Y}^\ast_{(\rho)}(\underline{x})\,
  &=&\delta(\lambda,\rho) \\
  \int \frac{d\mu(\lambda)}{(2\pi)^3} \,
  {\cal Y}_{(\lambda)}(\underline{x})\,
  {\cal Y}^\ast_{(\lambda)}(\underline{y})\,
  &=&\frac{\delta^3(\underline{x}-\underline{y})}{\sqrt{\s}}
\end{eqnarray}
where $\sqrt{\s}=f^2(\chi) \sin\theta$. The integration measures 
for the different types of Robertson-Walker metric are
\begin{eqnarray}
  \int d\mu(\lambda) = \left\{\begin{array}{lcc}
      \int_0^\infty dk \:\sum_{j=0}^\infty \:\sum_{m=-j}^j&&\kappa=-1, 0\\
       && \\
     \sum_{k=1}^\infty \:\sum_{j=0}^{k-1} \:\sum_{m=-j}^j&&\kappa=1
    \end{array}\right\}
\end{eqnarray}
which also specifies the form and range of the quantum numbers 
$k, j,$ and $m$.

\vspace{0.3cm}
For the spatially flat $\kappa=0$ metrics, it is of course 
frequently more convenient to choose the eigenfunctions 
in Cartesian form       
\be
  {\cal Y}_{(\lambda)}(\underline{x}) = \exp(i k_j x^j)
\ee
where $(\l) = (k_1,k_2,k_3)$ and $K^2(\l) = k^2$.
The measure is just $ \int d\mu(\lambda) = \int d^3\uk$.

\section{Green Functions}
\label{sec-green}

The Green function is given by the equation
\begin{equation}
  \sqrt{-g}(\Box + m^2 + \xi R){\cal G}(x,y) = - \delta^{(d+1)}(x - y)
\label{eq:greenfncst}
\end{equation}
As in the previous section, if the equation is translationally
invariant with respect to a particular coordinate then the Green
function may be expressed as a Fourier transform. 

In the static case,
\be
  {\cal G} (x,y) = \int_{-\infty}^{\infty} \frac{d\w}{(2\pi)} \, 
  e^{-i\w(x^0-y^0)} \tG
  (\w;\underline{x},\underline{y})
\ee
where
\be
  \sqrt{-g}(\Box_i -g^{00}\w^2 + m^2 + \xi R)\tG(\w;\underline{x},
  \underline{y})   = - \delta^d(\underline{x} - \underline{y})
\ee
For dynamic spacetimes,
\be
  {\cal G} (x,y) = \int \frac{d^d\underline{k}}{(2\pi)^d} \,
  e^{i\underline{k} . (\underline{x} - \underline{y})} \tG
  (\underline{k};x^0, y^0)
\ee
where
\be
  \sqrt{-g}(\Box_0 -g^{ij}k_ik_j + m^2 + \xi R)\tG(\underline{k};x^0,y^0) = -
  \delta(x^0 - y^0)
\ee
We cannot say more about the dynamic Green function in general.
However, in the static case, we can give a general solution. 
Using the orthonormality relations of solutions to the wave equation, 
we find 
\be
  {\cal G}(x,y)= \int^\infty_{-\infty} \frac{d\w}{(2\pi)} \, \int
  \frac{d\mu(\ll)}{(2\pi)^d} \, \frac{e^{-i\w(x_0-y_0)} \,
    \tpsi_{\bl}(\nu,\underline{x}) \,
    \tpsistar_{\bl}(\nu,\underline{y})}{(\w^2-\nu(\ll)^2)}
\ee
plus solutions of the homogeneous equation. The $\w$ integral is
along the real axis, but there are poles at $w=\pm \nu$. The choice of
contour ($i\e$ prescription) selects one of a variety of different Green
functions \cite{birrel82,fulling89}.

\begin{figure}[htb]
  \centerline{\epsfxsize=2.5in \epsfbox{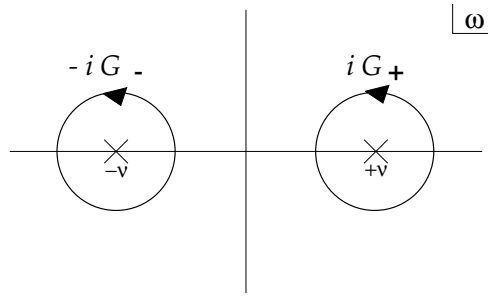}}
\caption{The omega contour integral for the Wightman functions.}
\label{fig:xfigwightman}
\end{figure}

For example, the contours giving the Wightman function are shown in
Fig(\ref{fig:xfigwightman}). Evaluating the integral gives
\begin{eqnarray}
  {\cal G}_+(x,y)&=& \langle 0 |\, \varphi(x) \, \varphi(y) \, 
  | 0 \rangle \\
  & & \nonumber \\
  &=&\frac{1}{2} \int \frac{d\mu(\ll)}{(2\pi)^d} \,
  \frac{1}{\w} \, e^{-i\w(x_0-y_0)} \, \tpsi_{\bl}(\w,\underline{x}) \,
  \tpsistar_{\bl}(\w,\underline{y})
\label{eq:wightman}
\end{eqnarray}
As the Wightman function is just the difference between two Green functions, 
it satisfies the homogeneous wave equation.

\section{Expectation values and two-point functions}
\label{sec-vev}

The vacuum expectation value of an operator $O(\varphi,\p)$ (a function of the 
field $\varphi$ and conjugate momentum $\pi$) in the Schr\"odinger picture is
\be
  \langle 0 | \, O(\varphi, \pi) \, | 0 \rangle \:=\: \int {\cal
    D}\phi \: \Psi^\ast_0 \: O(\phi, -i \frac{\delta}{\delta \phi}) \:
  \Psi_0
\ee
The action of the momentum is simply
\begin{eqnarray}
  \pi(\underline{x}) \Psi_0 &=& -i \frac{\delta}{\delta \phi
    (\underline{x})} \Psi_0 \nonumber \\ &=& i \int d^d\underline{y}
  \sqrt{ h_y \: h_x} \: \phi(\underline{y}) G(\underline{y},
  \underline{x};t)\: \Psi_0
\end{eqnarray}
To evaluate the action of the field, it is convenient to modify the wave functional
by formally introducing a `source', viz.
\be
  \Psi_J \:=\: \Psi_0 \:\exp{\bigl\{\half \int
    d^d\underline{x} \sqrt{-h_x} \:
    J(\underline{x})\,\phi(\underline{x}) \bigr\}}
\ee
Omitting field independent terms for clarity, the vacuum amplitude is then
\begin{eqnarray}
  \langle 0 | 0 \rangle_J &\sim& \int {\cal D}\phi e^{-\int d^d\underline{x} \int
    d^d\underline{y} \sqrt{h_x h_y}
    \phi(\underline{x})G_{\Re}(\underline{x},\underline{y};t)
    \phi(\underline{y}) \:+\: \int d^d\underline{x} \sqrt{-h_x}
    J(\underline{x}) \phi(\underline{x})} \\
    &\sim& \exp\Bigl\{\mbox{$\frac{1}{4}$} \int d^d\underline{x} \int
  d^d\underline{y} \sqrt{h_x h_y}
  J(\underline{x})\Delta_{\Re}(\underline{x},\underline{y};t)
  J(\underline{y})\Bigr\} 
\end{eqnarray}
Notice that the functional integral involves just the real part $G_{\Re}$
of the kernel. Also note that $\Delta_{\Re}$ is defined here as the inverse
of $G_{\Re}$ ({\it not}\/ the real part of the inverse kernel).
Expectation values of products of fields can then be found by repeated
differentiation with respect to the source, e.g.
\begin{eqnarray}
  \langle 0 |\varphi(\underline{x}) \varphi(\underline{y})| 0 \rangle &=& 
  \: \int {\cal D}\phi
  \: \Psi^\ast_J \: \phi(\underline{x}) \phi(\underline{y}) \: \Psi_J
  \Big|_{J=0}  \nonumber
  \\ &=& \frac{1}{\sqrt{-h_x}}\:
  \frac{\delta}{\delta J(\underline{x})} \:\:\frac{1}{\sqrt{-h_y}}\:
  \frac{\delta}{\delta J(\underline{y})} \;\; \langle 0 | 0 \rangle _J
  \Big|_{J=0} \nonumber \\ &=& \half
  \Delta_{\Re}(\underline{x},\underline{y};t)
\end{eqnarray}
This is the equal-time Wightman function, confirming the identifications
(\ref{eq:invwightmanmink}) and (\ref{eq:invwightmanstatic})
in the text.

Other useful two-point functions are
\begin{eqnarray}
  \langle 0 | \, \pi(\underline{x}) \, \pi(\underline{y}) \, | 0
  \rangle &=&
  \half \sqrt{h_x h_y} \Bigl\{
  \:G_{\Re}(\underline{x},\underline{y};t)
   \nonumber \\ &+&\! \!\!\!\! \int\! d^d\underline{u}
  \int\! d^d\underline{v}\, \sqrt{h_u h_v} \,
  G_{\Im}(\underline{x},\underline{u};t)
  \Delta_{\Re}(\underline{u},\underline{v};t)
  G_{\Im}(\underline{v},\underline{y};t)
  \Bigr\}
\end{eqnarray}
and
\begin{eqnarray}
  \langle 0 | \, [\varphi(\underline{x}) , \pi(\underline{y})] \, | 0
  \rangle &=& i \delta^d(\underline{x}-\underline{y})\\
  \langle 0 | \, \{\varphi(\underline{x}) , \pi(\underline{y})\} \, | 0
  \rangle &=&- \int d^d\underline{u} \, \sqrt{h_x h_u} \,
  G_{\Im}(\underline{x},\underline{u};t)
  \Delta_{\Re}(\underline{u},\underline{y};t)
\end{eqnarray}

\newpage
\baselineskip=14pt plus 2pt


\begin{thebibliography}{10}

\bibitem{birrel82}
N.~D. Birrel and P.~C.~W. Davies,
\newblock {Quantum Fields in Curved Space}
\newblock (Cambridge University Press, 1982).

\bibitem{fulling89}
S.~A. Fulling,
\newblock {Aspects of Quantum Field Theory in Curved Space-time}
\newblock (Cambridge University Press, 1989).

\bibitem{grib94}
A.~A. Grib, S.~G. Mamayev and V.~M. Mostepanenko,
\newblock {Vacuum Quantum Effects in Strong Fields}
\newblock (Friedmann Laboratory Publishing, St. Petersburg, 1994).

\bibitem{moss96}
I.~Moss,
\newblock {Quantum Theory, Black Holes and Inflation}
\newblock (John Wiley and Sons Ltd, 1996).

\bibitem{me96}
D.~V. Long and G.~M. Shore,
\newblock in preparation.

\bibitem{hawking75}
S.~W.~Hawking,
\newblock {Comm. Math. Phys.} 43 (1975) 199.

\bibitem{jackiw89}
R.~Jackiw,
\newblock in {Field Theory and Particle Physics}, $5^{th}$ Jorge Swieca 
Summer School, Brazil (World  Scientific, 1989).

\bibitem{schwinger51}
J.~Schwinger,
\newblock {Phys. Rev.} 82 (1951) 664.

\bibitem{dewitt65}
B.~S. DeWitt,
\newblock in {Relativity, Groups and
  Topology}, eds. B.~S. DeWitt and C.~DeWitt
 (Gordon and Breach, 1964).

\bibitem{kuchar76}
K.~Kucha\v{r},
\newblock {J. Math. Phys.} 17 (1976) 777, 792 and 801.

\bibitem{freese85}
K.~Freese, C.~T. Hill and M.~Mueller,
\newblock {Nucl. Phys.} B255 (1985) 693.

\bibitem{halliwell91}
J.~J. Halliwell,
\newblock {Phys. Rev.} D43 (1991) 2590.

\bibitem{guven89}
J.~Guven, B.~Lieberman and C.~T. Hill,
\newblock {Phys. Rev.} D39 (1989) 438.

\bibitem{eboli89}
O.~\'{E}boli, S.~Pi and M.~Samiullah,
\newblock {Ann. Phys.} 193 (1989) 102.

\bibitem{tod90}
L.~P.~Hughston and K.~P.~Tod,
\newblock {An Introduction to General Relativity}
\newblock (Cambridge University Press, 1990).

\bibitem{grad80}
I.~S. Gradshteyn and I.~M. Ryzhik,
\newblock {Table of Integrals, Series, and Products}
\newblock (Academic Press, New York, 1980).

\bibitem{magnus66}
W.~Magnus, F.~Oberhettinger and R.~P. Soni,
\newblock {Formulas and Theorems for Special Functions of Mathematical
  Physics}
\newblock (Springer-Verlag, Berlin, 1966).

\bibitem{bernard77}
C.~Bernard and A.~Duncan,
\newblock {Ann. Phys.} 107 (1977) 201.

\bibitem{fulling73}
S.~A. Fulling,
\newblock {Phys. Rev.} D7 (1973) 2850.

\bibitem{parker74}
L.~Parker and S.~A. Fulling,
\newblock {Phys. Rev.} D9 (1974) 341.

\bibitem{lifshitz63}
E.~M. Lifshitz and I.~M. Khalatnikov,
\newblock {Adv. Phys.} 12 (1963) 185.

\bibitem{ford76}
L.~H. Ford,
\newblock {Phys. Rev.} D14 (1976) 3304.

\end{thebibliography}
\end{document}